\documentclass[aps,
		prd,
		reprint,
		twocolumn,
		superscriptaddress,
		shortbibliography,
		nofootinbib,
		floatfix,
		notitlepage
		]{revtex4-2}
\pdfoutput=1

\usepackage{amsmath,amssymb}
\usepackage{graphicx,accents}
\usepackage{graphicx,epstopdf}
\usepackage[usenames,dvipsnames]{xcolor}
\usepackage{slashed}
\usepackage[normalem]{ulem}
\usepackage{bm}
\usepackage{bbm}
\usepackage{bbold}
\usepackage[dvipsnames]{xcolor}
\usepackage{hyperref}
\hypersetup{colorlinks=true}
\usepackage{tabularx}

\newcommand{\vphi}{\varphi}
\newcommand{\eps}{\varepsilon}

\newcommand{\mbf}[1]{\mathbf{#1}}
\newcommand{\trm}[1]{\textrm{#1}}
\newcommand{\tsf}[1]{\textsf{#1}}

\newcommand{\be}{\begin{equation}}
\newcommand{\ee}{\end{equation}}
\newcommand{\bea}{\begin{eqnarray}}
\newcommand{\eea}{\end{eqnarray}}
\newcommand{\bi}{\begin{itemize}}
\newcommand{\ei}{\end{itemize}}

\newcommand{\pdeg}{\Gamma}

\newcommand{\nn}{\nonumber}

\newcommand{\prob}{\tsf{P}}

\newcommand{\cpo}{\tiny\tsf{cp}}
\newcommand{\lpo}{\tiny\tsf{lp}}
\newcommand{\LCperp}{{\scriptscriptstyle \perp}}
\newcommand{\LCpara}{{\scriptscriptstyle \parallel}}

\newcommand{\figref}[1]{Fig. \ref{#1}}

\newcommand{\eqnref}[1]{Eq. (\ref{#1})}

\newcommand{\appref}[1]{App. \ref{#1}}

\newcommand{\vkap}{\varkappa}

\newcommand{\ud}{\mathrm{d}}

\makeatletter
\def\ps@pprintTitle{%
 \let\@oddhead\@empty
 \let\@evenhead\@empty
 \def\@oddfoot{}%
 \let\@evenfoot\@oddfoot}
\makeatother

\begin{document}

\title{Locally monochromatic two-step nonlinear trident process in a plane wave}

\author{S.~Tang}
\affiliation{College of Physics and Optoelectronic Engineering, Ocean University of China, Qingdao, Shandong, 266100, China}

\author{B.~King}
\email{b.king@plymouth.ac.uk}
\affiliation{Deutsches Elektronen-Synchrotron DESY, Notkestr. 85, 22607 Hamburg, Germany}
\affiliation{Centre for Mathematical Sciences, University of Plymouth, Plymouth, PL4 8AA, United
Kingdom}


\date{\today}
\begin{abstract}
In many-cycle plane waves at intermediate intensities, the nonlinear trident process can be well-approximated by the two sequential steps of nonlinear Compton scattering of a polarised real photon followed by its transformation into an electron-positron pair via nonlinear Breit-Wheeler pair creation. We investigate this two-step process in the intermediate intensity regime by employing the locally monochromatic approximation for each step and numerically evaluating resulting expressions. When photon polarisation is included, it is found to produce an order $10\%$ decrease in the trident rate: the importance of polarisation increases at lower intensities, and decreases at higher intensities. Its importance persists at higher intensities in a linearly-polarised background, but disappears at high intensities in a circularly-polarised background. If the two steps are made to take place in two linearly-polarised plane wave pulses with perpendicular polarisations, the pair yield can be increased by approximately $30\%$ compared to two plane waves with the same polarisation. It is also shown that harmonic structures in the Compton step can be passed to the pair step if the Compton edge is at an energy of the order of the threshold for linear Breit-Wheeler.
\end{abstract}
\maketitle
\twocolumngrid

\section{Introduction}

In the landmark E144 experiment at SLAC in the mid 1990s, a $46.6\,\trm{GeV}$ electron beam was collided with a weakly-focussed laser of moderate intensity parameter ($\xi \approx 0.3$) in a near head-on collision \cite{E144:1996enr,burke97,bamber99}. This provided the first measurement of nonlinear Compton scattering (NLC) of real high energy photons and their transformation into electron-positron pairs via the nonlinear Breit-Wheeler (NBW) process in the multiphoton regime. This sequence of processes is the `two-step' part of the nonlinear trident process $e^{\pm}\to e^{\pm}+e^{+}e^{-}$. Recently, the NA63 experiment at CERN measured the two-step \emph{nonlinear} trident process with $\xi \sim O(10)$, at strong-field parameter $\chi\approx2.4$ in the collision of $200\,\trm{GeV}$ electrons with germanium single crystals \cite{Nielsen:2022bws}. It is planned, at the E320 experiment at SLAC \cite{chen22} and the LUXE experiment  at DESY \cite{Abramowicz:2021zja}, to measure this two-step process in the intermediate intensity regime $\xi\sim\mathcal{O}(1)$. Since the intensity parameter represents the charge-field coupling, these new experiments will probe strong-field quantum electrodynamics (QED) effects where the charge-field interaction is non-perturbative and can no longer be described using leading-order multiphoton contributions.

Although exact solutions for the complete nonlinear trident process have been calculated in a plane wave pulse \cite{hu10,ilderton11,Hu:2013yrz, Dinu:2017uoj,Mackenroth:2018smh,Dinu:2019wdw,Torgrimsson:2022ndq} also including radiation reaction \cite{Torgrimsson:2022ndq} (as well as in a constant crossed field \cite{baier72,ritus72,morozov77,king13b,king18c}, a magnetic field \cite{Novak:2012zz}, a Coulomb field \cite{Torgrimsson:2020wlz} and also studied using an adiabatic approximation \cite{Titov:2021kbj} and in trains of laser pulses \cite{Kaminski:2022uoi}), they are computationally expensive and a more efficient and versatile method is required to model experiments. 
Therefore, a simulational approach that calculates particle trajectories classically, and adds first-order QED effects such as the NLC and NBW processes  via Monte-Carlo methods, is typically favoured in designing and analysing experiments \cite{bamber99, cole18,poder18,DiPiazza:2019vwb,2021NJPh...23j5002S,chen22,Abramowicz:2021zja}. For second (and higher) order processes, the correct factorisation of the $n$-step subprocess requires taking into account the polarisation of intermediate particles. 
For the nonlinear trident process, this means the polarisation of the intermediate photon must be included in the NLC and NBW steps~\cite{ritus72}. The inclusion of photon polarised strong-field QED processes into simulational approaches is a relatively recent development \cite{king13a,PhysRevLett.124.014801,PhysRevResearch.2.032049,Seipt-2021,Seipt-2022,Nielsen:2022bws} and is so far achieved using the locally constant field approximation (LCFA)~\cite{ritus85,Seipt:2020diz,Fedotov:2022ely}. However,  at intermediate intensity values $\xi \sim O(1)$, the LCFA for the first-order processes is known to become inaccurate for typical experimental parameters \cite{harvey15,DiPiazza:2017raw,Ilderton:2018nws,DiPiazza:2018bfu,King:2019igt}.

In the current paper, we assess the importance of photon polarisation in modelling the two-step nonlinear trident process. 
This is achieved by adapting the locally monochromatic approximation (LMA) \cite{Heinzl:2020ynb,Blackburn:2021cuq,Blackburn:2021rqm}, which is being used to model the interaction-point physics of the LUXE experiment \cite{Abramowicz:2021zja} as well as strong-field QED experiments at the BELLA PW laser \cite{Turner:2022hch} (also appearing in other forms \cite{Yokoya:1991qz,Hartin:2018egj} and being used in the modelling of the E144 experiment \cite{bamber99}), to include photon-polarised rates for the NLC and NBW processes. Photon polarisation has been studied in NLC in monochromatic backgrounds \cite{Ivanov:2004fi} and in plane-wave pulses \cite{TangPRA022809,TANG2020135701}; photon polarisation in NBW in the LMA and in plane-wave backgrounds has also been studied \cite{Tang:2022a,TangPRD056003}; here we gather and present the LMA formulas in a circularly-polarised and linearly-polarised background.
A second focus of this paper is to consider the role of harmonics: whether the harmonic structure of nonlinear Compton, which is a key experimental observable, e.g. in LUXE, can be found in the pair spectrum in the nonlinear trident process. 
The LMA was chosen as it has been shown that first-order subprocesses can be described more accurately at intermediate intensity values, which are planned to be used in upcoming experiments \cite{Blackburn:2021cuq,Blackburn:2021rqm}.
Although it is known that there are effects missed by the LMA that are related to the finite bandwidth of the background pulse, such as harmonic broadening, low-energy photons produced in  NLC~\cite{King:2020hsk} and pairs created by the low intensity part of the pulse \cite{Tang:2021qht}, these should be negligible for realistic set-ups with the parameters we consider here. (For more background on strong-field QED, we direct the reader to the reviews \cite{ritus85,dipiazza12,Fedotov:2022ely,RMP2022_045001}.)

Overall, trident is a $1\to 3$ process: $e^{\pm} \to e^{\pm} + e^{+}e^{-}$, but we are also interested in the intermediate steps, which are both $1\to 2$ processes: $e^{\pm} \to e^{\pm} + \gamma$ followed by $\gamma \to e^{+}e^{-}$, and in the two-step case, we take the photon to be on-shell. In this paper, we assume the initial particle is an electron (analogous conclusions follow for a positron).  The total probablity will depend on: i) the incoming particle's energy parameter $\eta = \vkap \cdot p / m^{2}$, where $\vkap$ is the plane wave background wavevector, $p$ is the probe electron momentum, $m$ is the mass of the electron; ii) the intensity parameter $\xi$ of the background, which will be defined in the following in terms of the gauge potential and iii) the number of cycles, $N$ of the background. We mainly focus on the total probability, $\prob = \prob(\xi,\eta,N)$ and lightfront momentum spectrum of the two-step trident process. The parameter space is large and we will also be studying the impact of photon polarisation on the total probability. Therefore, in this paper we will restrict our attention to the most relevant parameter ranges, given below.

The initial particle energy will mainly be chosen to be $16.5\,\trm{GeV}$ and the laser photon energy to be $1.55\,\trm{eV}$ ($800\,\trm{nm}$ wavelength), or its third harmonic $4.65\,\trm{eV}$. This particle energy is chosen as it is the energy of electrons planned to be used at LUXE, and is similar to the energy of $13\,\trm{GeV}$ used at E320. To investigate higher harmonics, we will calculate one example with $80\,\trm{GeV}$ and $500\,\trm{GeV}$ electrons, which are both currently much higher than planned for in experiments.
The investigation into potential harmonic structure is particularly relevant for our approach using the LMA, where such effects are captured, in comparison to approaches based on the LCFA, which do not capture harmonic effects. 
For convenience, in this paper we will use a cosine-squared pulse envelope. We note that this pulse shape has a much wider bandwidth than can be transmitted through optical elements in an experiment. As a consequence, when $\xi\lesssim 1$ and initial electron energies are lower than around $10\,\trm{GeV}$, an exact plane-wave calculation will show bandwidth effects beyond the LMA can become dominant. However, these effects will not be present in an experiment. For consistency, we consider initial electron energies higher than this. (See \appref{app:lowEnergy} for more details on this point.)

The intensity parameter will be varied in the `intermediate regime', $\xi\sim O(1)$, which we take to be $\xi = 0.5 ... 5$. This is significant because the LMA is accurate in this regime, whereas studies of trident based on the LCFA, are limited to higher values of $\xi$. This parameter regime will also be probed by the E320 and LUXE experiments.

The number of laser cycles $N$ is a less important parameter in the LMA because it can be factored out and occurs just as a pre-integral over the laser phase. However, $N$ must be large enough that the LMA is a good approximation to the full QED probability (it is assumed that $N\gg 1$ for this to be the case). We will consider a gauge potential $a=|e|A$ describing the background as a plane wave pulse. The potential depends upon the phase $\phi=\vkap \cdot x$ in the form:
\begin{equation}
\mbf{a}^{\perp} = \begin{cases}
m\xi f(\phi) (\cos \phi, \varsigma\sin\phi ) & |\phi| < \Phi/2 \\
0 & \trm{otherwise},\label{eqn:adef}
\end{cases} 
\end{equation}
where $a = (0, \mbf{a}^{\perp},0)$ and $\varsigma \in \{-1,0,1\}$ is chosen to switch between linear ($\varsigma=0$) or circular ($\varsigma=\pm 1$) polarisation (we note that the time-averaged amplitude of the potential is different for these two choices).  
In this paper, we will always use $f(\phi) = \cos^{2}(\pi\phi/\Phi)$ and $\varsigma=1$ for circular polarisation, where $\Phi = 2\pi N$, and will choose $N=16$, which corresponds to a full-width-at-half-maximum duration of $21.3\,\trm{fs}$ for a wavelength of $\lambda = 2\pi/\vkap^{0}=800\,\trm{nm}$.

Let us denote the lightfront momentum fraction of the photon $s_{\gamma}=\vkap \cdot \ell/\vkap\cdot p$, where $\ell$ is photon momentum. The probability for nonlinear Compton scattering of a photon in polarisation state $\sigma$ is $\tsf{P}_{\gamma}^{\sigma}$, and nonlinear Breit-Wheeler pair-creation from a photon in the same state is $\tsf{P}_{e}^{\sigma}$. Then the total probablity $\tsf{P}$ for the two-step trident process, can be written \cite{ritus72}:
\bea
\tsf{P} = \sum_{\sigma}\int_{0}^{1} ds_{\gamma} \int d\phi\, \frac{d^{2}\tsf{P}^{\sigma}_{\gamma}(\phi; s_{\gamma})}{ds_{\gamma}\,d\phi}~\tsf{P}^{\sigma}_{e}(\phi; s_{\gamma}),\label{eqn:tri2}
\eea
where $\tsf{P}^{\sigma}_{e}(\phi; s_{\gamma}) = -\int_{\phi} \,d\phi'~ d\tsf{P}^{\sigma}_{e}(\phi';s_{\gamma})/d\phi'$ and ${\sigma\in\{\parallel,\perp\}}$
refers to the two polarisation eigenstates of the background. 
For a linearly-polarised background, the designation $\parallel$ ($\perp$) refers to the photon polarisation eigenstate being in the same (opposite)
polarisation state as the background with a fixed direction in lab co-ordinates. For a circularly-polarised background, these directions rotate with $\parallel$ ($\perp$) referring to when a photon collides head-on with the background having a polarisation that rotates in the same (opposite) direction as the background\footnote{We note that in terms of helicity states, $\parallel$ ($\perp$) refer to the photon have the opposite (same) helicity as the background.}.
(The polarised NBW probability $\prob^{\sigma}_{e}$ is defined such that the total unpolarised probability, $\prob_{e} = \prob^{\parallel}_{e} + \prob^{\perp}_{e}$ i.e. the usual $1/2$ polarisation averaging factor \cite{ritus72} is included already in $\prob^{\sigma}_{e}$, and analogously for $\prob_{\gamma}^{\sigma}$.) In order to assess the importance of photon polarisation, we will also be interested in the approximation to this probability that uses unpolarised probabilities, $\tsf{P}_{\gamma}$ and $\tsf{P}_{e}$ for the nonlinear Compton and Breit-Wheeler steps respectively. We denote:
\bea
\widetilde{\tsf{P}} = \int_{0}^{1} ds_{\gamma} \int d\phi\, \frac{d^{2}\tsf{P}_{\gamma}(\phi; s_{\gamma})}{ds_{\gamma}\,d\phi}~\tsf{P}_{e}(\phi; s_{\gamma}).
\eea
(This approximation to the two-step trident probability has often been used in numerical simulations of trident and QED cascades \cite{PRL2008200403,Kirk_2009,nerush11,elkina11,TangPRA2014,gonoskov17,Samsonov:2018nff,Khudik:2018hkr}.)
Since the total probability, \mbox{\eqnref{eqn:tri2}}, involves a sum over polarisation eigenstates of the background, to assess the importance of photon polarisation, one should consider different background polarisation. Here, we study the cases of a circularly- and a linearly-polarised plane wave pulse.

Probabilities for sub-processes will be calculated within the locally monochromatic approximation \cite{Heinzl:2020ynb,Tang:2022a}. This is an adiabatic approximation that neglects derivatives of the slow timescale of the pulse envelope, but includes the fast timescale of the carrier frequency exactly. It can be shown to be equal to the leading order term in an expansion of the total probability in $\Phi^{-1}$ \cite{Torgrimsson:2020gws}.

This paper is organised as follows. In Sec. II the results are presented for the two-step nonlinear trident process in a circularly-polarised (Sec. II A) and a linearly-polarised (Sec. II B) background. The results are discussed in Sec. III and conclusions drawn in Sec. IV. In Appendix A, the LMA formulas for photon-polarised nonlinear Compton scattering and photon-polarised nonlinear Breit-Wheeler are given and Appendix B contains further details about bandwidth effects in the LMA.

\section{LMA for two-step trident process}
In this section, we combine the LMA formulas (shown in \appref{AppA}) for the NLC and NBW processes~\cite{Heinzl:2020ynb,Tang:2022a} to calculate the yield of the two-step trident process.
The lightfront momentum is conserved in the interaction with the plane-wave. We use the variable $0 \leq s \leq 1$ to denote what fraction a particle's lightfront momentum is of the initial electron lightfront momentum. This can be written:
\be 
1 = s_{e'} + s_{\gamma}; \quad s_{\gamma} = s_{q}+s_{e''}, \label{eqn:sdef}
\ee
where $s_{\gamma}$ is the lightfront momentum fraction of the photon, $s_{e'}$ of the scattered electron, $s_{e''}$ of the created electron and $s_{q}$ of the positron. In this paper, only $s_{\gamma}$ and $s_{q}$ will appear explicitly; the other lightfront momentum fractions will be integrated out. Then we can write the total probability as an integral over the double differential probability:
\be\prob =\int^{1}_{0}\ud s_{q}\int^{1}_{s_{q}}  \ud s_{\gamma} ~ \frac{\ud^{2}\prob}{\ud s_{q}\ud s_{\gamma}}\,, \ee
where the double differential involves two integrals over the average phase positions of incoming particle in NLC and NBW:
\be 
\frac{\ud^{2}\prob}{\ud s_{q}\ud s_{\gamma}} = \sum_{\sigma}\int^{\phi_{f}}_{\phi_{i}} \ud\phi \frac{\ud^{2}\prob^{\sigma}_{\gamma}(\phi; s_{\gamma})}{\ud s_{\gamma}\ud\phi} \int^{\phi_{f}}_{\phi} \ud\phi' \frac{\ud^{2}\prob^{\sigma}_{e}(\phi';s_{q})}{\ud s_{q}\ud\phi'}
\label{Eq-QED-trident}
\ee 
(where again $\sigma \in \{\parallel,\perp\}$). The production of the positron depends not only on the energy of the NLC photon, but also on the polarisation of the photon.
The polarisation degree of the NLC photon is defined as
\be
\Gamma(s_{\gamma})=\frac{\ud \prob^{\LCpara}_{\gamma}/\ud s_{\gamma} - \ud \prob^{\LCperp}_{\gamma}/\ud s_{\gamma}}{\ud \prob_{\gamma}/\ud s_{\gamma}}
\label{eqn:GammasDef}
\ee
where $\ud \prob_{\gamma}/\ud s_{\gamma}=\ud \prob^{\LCpara}_{\gamma}/\ud s_{\gamma} + \ud \prob^{\LCperp}_{\gamma}/\ud s_{\gamma}$. Following the definition in \eqnref{eqn:GammasDef}, we define the polarisation degree for a distribution of photons to be $\Gamma$ where  $\Gamma= (\prob^{\LCpara}_{\gamma}-\prob^{\LCperp}_{\gamma})/\prob_{\gamma}$ and $\prob_{\gamma} = \prob^{\LCpara}_{\gamma}-\prob^{\LCperp}_{\gamma}$. For photons in the $\parallel$ polarisation eigenstate, $\Gamma=1$; if they are in the $\perp$ eigenstate then $\Gamma=-1$ and if they are completely unpolarised, $\Gamma=0$.

\subsection{Circularly polarised background}
First, we investigate the potential harmonic structure in the energy spectrum of the outgoing positron. 
For the harmonic structure from both NLC and NBW steps to be discernible: i) pairs must be produced by photons with momentum fractions around the first NLC harmonic $s_{\gamma,1} = 2\eta/(2\eta +1 + \xi^2)$; ii) the photon momentum fraction must be larger than the threshold for pair creation from a single laser photon, i.e. linear Breit Wheeler from the carrier frequency, which requires $2(1+\xi^{2})/(\eta s_{\gamma,1})<1$. Combining these conditions, one arrives at the requirement on the initial electron's energy parameter of $\eta>(1+\sqrt{2})(1+\xi^2)$. For $\xi=1$, and a standard laser carrier frequency of $1.55\,\trm{eV}$, this corresponds already to $407\,\trm{GeV}$ electrons (or an energy parameter of $\eta=4.83$). 
Therefore to demonstrate harmonic structure in the pair spectrum, we consider $500\,\trm{GeV}$ electrons colliding head-on with a $16$ cycle, $\xi=1$ circularly-polarised plane wave with carrier frequency $1.55\,\trm{eV}$  (i.e. $\eta=5.94)$.

\begin{figure}[h!!]
\includegraphics[width=8cm]{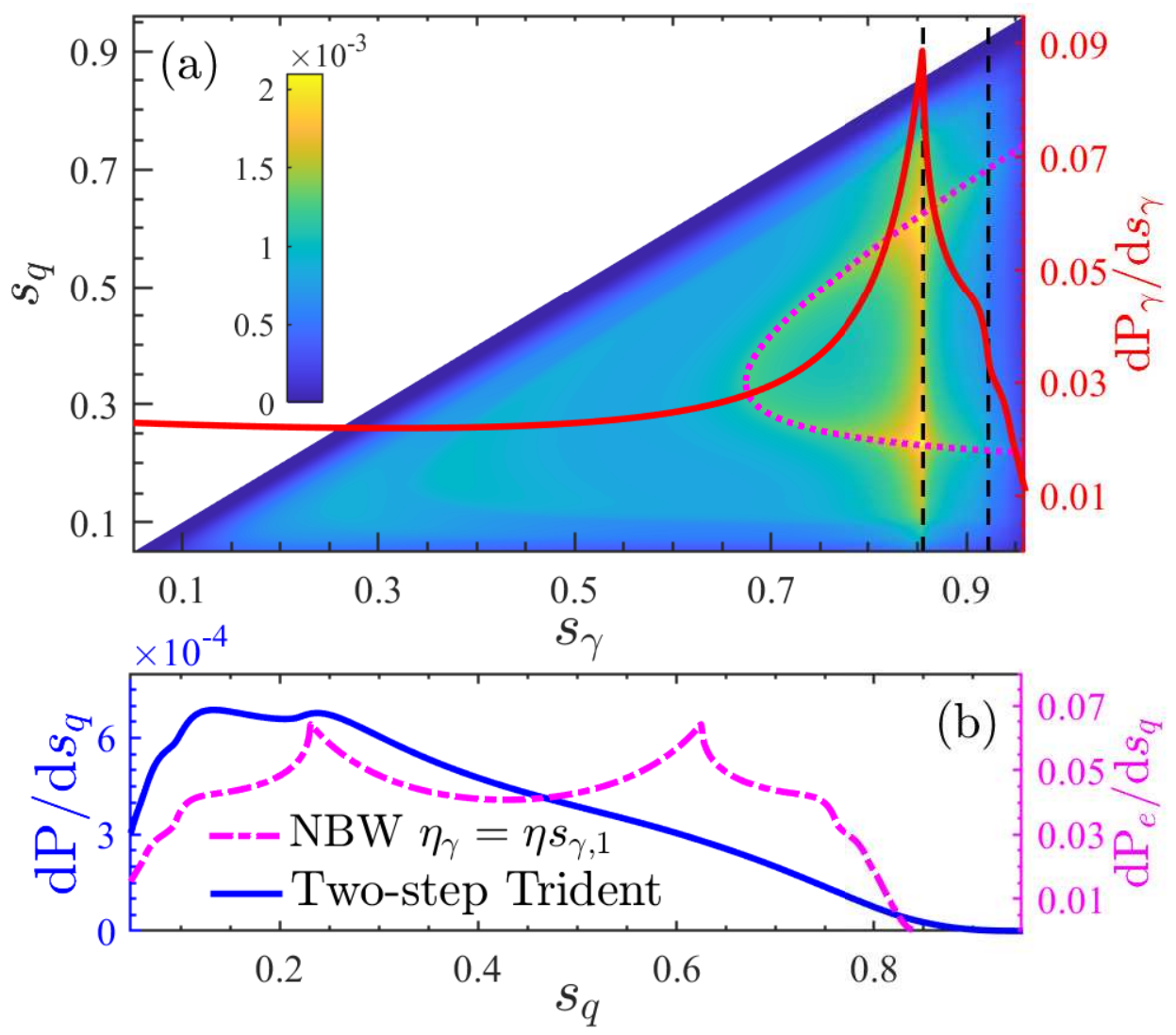}
\caption{Differential probability for the two-step trident process in a $\xi=1$, $N=16$, circularly polarised pulse of frequency $1.55\,\trm{eV}$ colliding head on with a $500\,\trm{GeV}$ electron, corresponding to an energy parameter of $\eta= 5.94$.
(a) Double lightfront spectrum, $\ud^{2}\prob/\ud s_{q}/\ud s_{\gamma}$ is plotted for the polarised intermediate photon and compared to the NLC spectrum (red overlaid line). The black dashed lines give the location of the first two harmonics in the photon spectrum at $s_{\gamma,n}=2n\eta /(2n\eta + 1 + \xi^2)$ for $n=1,2$; the magenta dotted line denotes the location of the first harmonic in the positron spectrum created by the photon with the energy $\eta s_{\gamma}$ at $s_{q} = \{1\pm[1-2(1+\xi^2)/(s_{\gamma}\eta)]^{1/2}\}/2$.
(b) Positron energy spectrum for the two-step trident process $\ud\trm{P}/\ud s_{q}$ and for the NBW process $\ud\trm{P}_{e}/\ud s_{q}$ created by the first-harmonic photon (with the energy parameter $\eta_{\gamma}=\eta s_{\gamma,1}$) in the NLC spectrum.
}
\label{Fig-Tri-E500-H1-Dist}
\end{figure}
The double differential lightfront spectrum, $\ud^{2}\prob/\ud s_{q}/\ud s_{\gamma}$ for this case is plotted in Fig.~\ref{Fig-Tri-E500-H1-Dist} (a).
We see the main features: i) the harmonic structures from the NLC spectrum are observed around the black dashed lines, namely $s_{\gamma,n}=2n\eta/(2n\eta + 1 + \xi^2)$ for $n=1,2$. This comparison is made manifest by the red solid line which plots the NLC spectrum of photons before the pair is created;
ii) the double differential spectrum rises up around the magenta dotted line, $s_{q} = \{1\pm[1-2(1+\xi^2)/(s_{\gamma}\eta)]^{1/2}\}/2$, corresponding to the edge of the first NBW harmonic triggered by the intermediate photon with the momentum fraction $s_{\gamma}>2(1+\xi^2)/\eta$.
This is illustrated by the location of harmonic peaks in the NBW positron spectrum [magenta dot-dashed line plotted in \mbox{Fig.~\ref{Fig-Tri-E500-H1-Dist} (b)}] created by a photon with energy equal to the edge of the first NLC harmonic range (the `Compton edge' \cite{Harvey09}), with energy $\eta_{\gamma}=\eta s_{\gamma,1}$.
The double differential spectrum peaks up around the crossing points between the first NLC-harmonic (black dashed) line and the first NBW-harmonic (magenta dotted) line in Fig.~\ref{Fig-Tri-E500-H1-Dist} (a).
After integrating over the photon lightfront momentum $s_{\gamma}$, the lower harmonic peak ($s_{q}=0.23$) remains in the positron spectrum of the nonlinear two-step trident process as shown in Fig.~\ref{Fig-Tri-E500-H1-Dist} (b) (but the upper harmonic peak around $s_{q}=0.62$ is smoothed out.)
\begin{figure}[h!!]
\includegraphics[width=8cm]{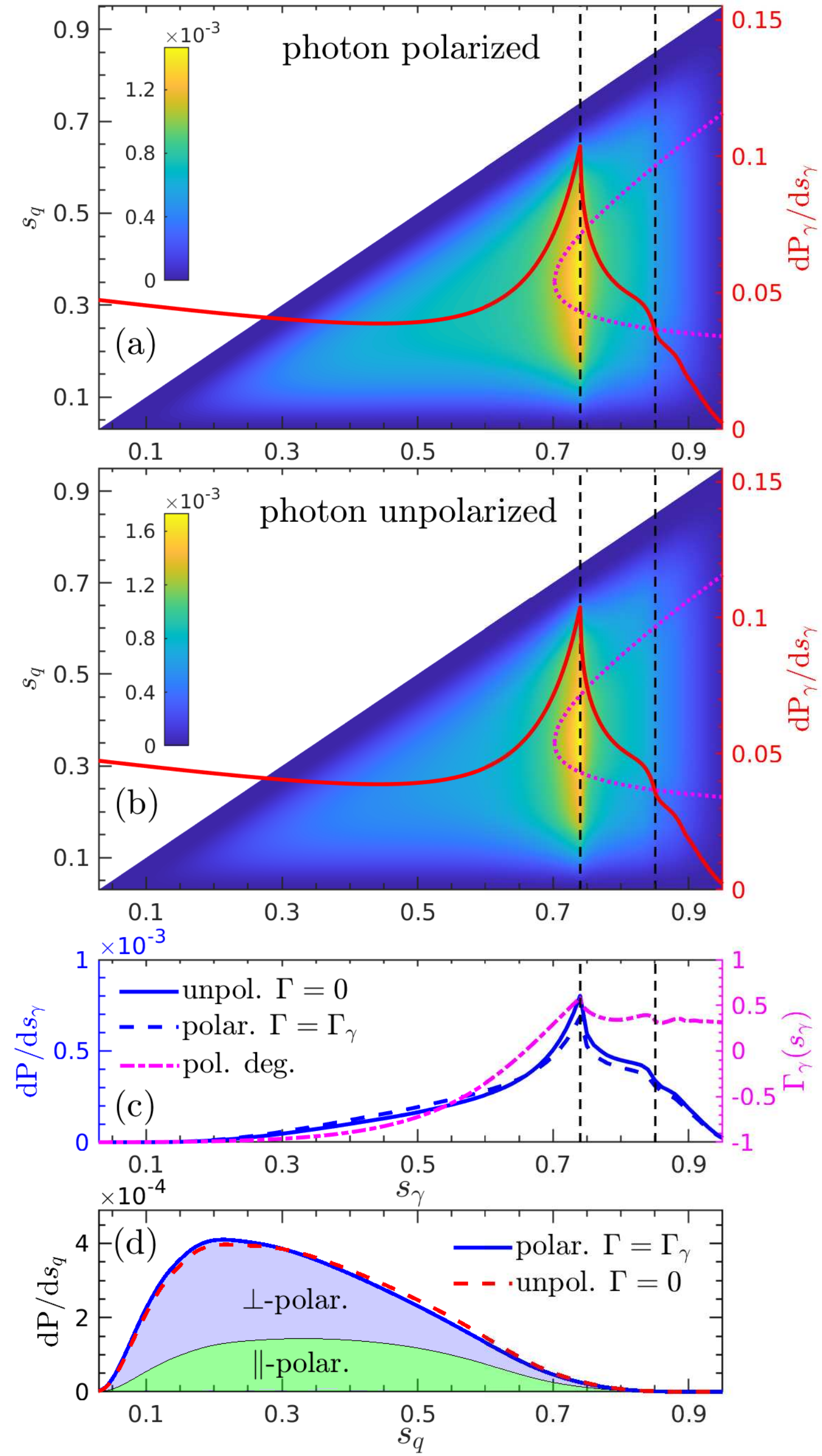}
\caption{Differential probability for the two-step trident process in a $\xi=1$, $N=16$, circularly polarised pulse of frequency $4.65\,\trm{eV}$ colliding head on with a $80\,\trm{GeV}$ electron, corresponding to an energy parameter of $\eta=2.85$. The double lightfront spectra $\ud^{2}\prob/\ud s_{q}/\ud s_{\gamma}$ lor an (a) polarised and (b) unpolarised photon and compared to the NLC spectrum (red overlaid line). The magenta dotted lines denote the location of the leading-NBW harmonic in the positron spectrum created by the photon with energy $\eta s_{\gamma}$ at $s_{q} = \{1\pm[1-2(1+\xi^2)/(2s_{\gamma}\eta)]^{1/2}\}/2$. (c) Dependence of the probability on the lightfront momentum of a polarised (blue dashed line) and unpolarised (blue sold line) photon: for the polarised case, the polarisation degree  is given by the magenta dashed-dotted line.
(d) Energy spectrum of the positron, $\ud\trm{P}/\ud s_{q}$, for unpolarised ($\Gamma=0$) and polarised ($\Gamma=\Gamma_{\gamma}$) intermediate photons. The shaded areas underneath the photon-polarised spectrum denote the contribution from each polarisation eigenstate.
 In (a)-(c), the black dashed lines give the location of the first two harmonics in the photon spectrum at $s_{\gamma}=2n\eta /(2n\eta + 1 + \xi^2)$ for $n=1,2$.}
\label{Fig-Tri-E80-H3-Dist}
\end{figure}

In Figs.~\ref{Fig-Tri-E80-H3-Dist}\,(a) and (b), the double differential spectrum for a lower-energy case is plotted, in which an $80\,\trm{GeV}$ electron collides head-on with a $16$ cycle circularly polarised plane wave pulse with carrier frequency at the third harmonic of the laser at $4.65\,\trm{eV}$, corresponding to an energy parameter of $\eta=2.85$. 
In this case, harmonic structure is again present in the NLC photon spectrum (illustrated by the red solid line), but the energy parameter is significantly lower, $\eta<2(1+\xi^2)$, so that pairs are only created via the nonlinear process, i.e. requiring higher harmonics from the laser. 
The leading NBW harmonic is now the second order with smooth harmonic edges, as shown in the figure by the magenta dotted line plotted at $s_{q} = \{1\pm[1-2(1+\xi^2)/(2s_{\gamma}\eta)]^{1/2}\}/2$.
Therefore, there is no noticeable NBW-harmonic structure remaining in the double differential spectrum of the trident process.

\begin{figure}[t!!!!]
 \center{\includegraphics[width=0.45\textwidth]{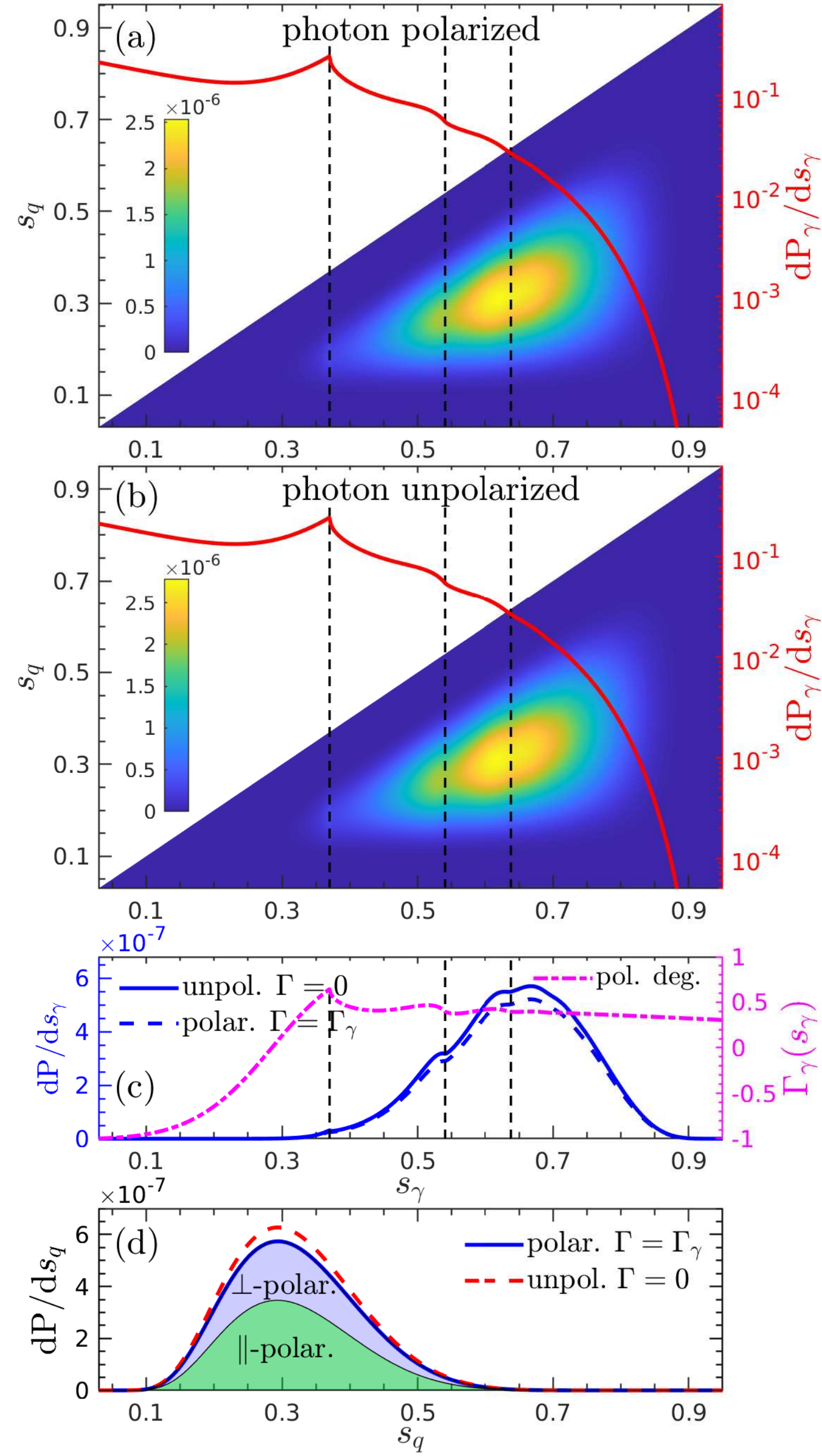}
\caption{Differential probability for the two-step trident process in a $\xi=1$, $N=16$, circularly polarised pulse of frequency $4.65\,\trm{eV}$ colliding head on with a $16.5\,\trm{GeV}$ electron, corresponding to an energy parameter of $\eta= 0.59$. 
The double lightfront spectra, $\ud^{2}\trm{P}/\ud s_{q}/\ud s_{\gamma}$ are plotted for an (a) polarised and (b) unpolarised photons and compared to the NLC spectrum (red overlaid line).
(c) The dependence of the probability on the lightfront momentum of a polarised (blue dashed line) and unpolarised (blue sold line) photon is plotted: for the polarised case, the polarisation degree is given by the magenta dashed-dotted line.
(d) The energy spectrum of the positron, $\ud\trm{P}/\ud s_{q}$, for unpolarised ($\Gamma=0$) and polarised ($\Gamma=\Gamma_{\gamma}$) intermediate photon is plotted. The shaded areas underneath the polarised spectrum denote the contributions from each polarisation eigenstate. In (a)-(c), the black dashed lines give the location of the first three harmonics in the photon spectrum at $s_{\gamma}=2n\eta /(2n\eta + 1 + \xi^2)$ for $n=1,2,3$.}}
\label{Fig-Tri-E165-H3-Dist}
\end{figure}

In Fig.~\ref{Fig-Tri-E80-H3-Dist}\,(c), the single differential spectrum $\ud\prob/\ud s_{\gamma}$ is plotted, which shows the spectrum of photons that created pairs in the two-step trident process. The locations of the Compton harmonics can be clearly seen. We note that the major contribution originates from the middle of the $s_{\gamma}$-range.
There is clearly a competition between: i) more photons being produced at low values of $s_{\gamma}$ and ii) pairs being created more easily for photons with higher values of $s_{\gamma}$.
The optimum region of $s_{\gamma}$ falls between these two extremes.
For the parameters in Fig.~\ref{Fig-Tri-E80-H3-Dist}, the low-order harmonics in the photon spectrum coincide with the optimal region, and therefore clear harmonic structures from the NLC spectrum can be observed in the double differential plots in Figs.~\ref{Fig-Tri-E80-H3-Dist} (a) and (b).
(As we will see later in Fig.~\ref{Fig-Tri-E165-H3-Dist}, if the initial electron energy is lowered to $16.5\,\trm{GeV}$, this optimal region of the photon spectrum corresponds to higher harmonic order and hence no harmonic structure is observable.)
Despite the appearance of harmonic structure in the double differential for the $80\,\trm{GeV}$ electron case due to NLC, the harmonic structure is not passed on to the energy spectrum of the produced pair. This is because: i) the Compton spectrum is integrated over and ii) pairs can only be created at higher harmonic order. 

Now we consider the effect of photon polarisation. For some parameters, the photon from NLC can be highly polarised \cite{TangPRA022809,TANG2020135701}, and since NBW can be strongly affected by the polarisation of the photon \cite{Toll:1952rq,Seipt:2020diz,Tang:2022a,TangPRD056003}, we expect photon polarisation to play a role in the two-step trident process. The dependence of the differential probability on the polarisation of the photon is plotted in Fig.~\ref{Fig-Tri-E80-H3-Dist}. First, we see that the harmonic position in the double differential spectrum is unchanged when the photon is unpolarised in Fig.~\ref{Fig-Tri-E80-H3-Dist} (b), which is to be expected, since polarisation can affect energy spectra, but not kinematic ranges. However, in Fig.~\ref{Fig-Tri-E80-H3-Dist} (c), we see that the \emph{height} of the harmonic peak is reduced. This has a straightforward explanation when compared to the photon polarisation degree at the harmonic peak. At this energy, photons are \emph{more} likely to be produced by NLC in the $\parallel$ polarisation state, which is the state \emph{least} likely to produce pairs via NBW. This is why, for these parameters, including photon polarisation leads to a suppression of pairs at the leading Compton harmonic. In general, we can see that photon polarisation can have the affect of enhancing and suppressing different parts of the pair spectrum. For example in the same plot, we see that at photon energies lower than the first Compton harmonic, the spectrum of pairs is instead \emph{enhanced}. Again, this can be understood by comparing with the polarisation degree, where we see that NLC is more likely to produce photons polarised in the $\perp$ polarisation state, which is \emph{more} likely to lead to NBW pair creation. Whilst photon polarisation can influence the pair spectrum, the total yield is not always affected: in Fig.~\ref{Fig-Tri-E80-H3-Dist} (d) it can be seen that the integrals of the polarised and unpolarised spectra are approximately equal. 
However, for higher energies, e.g. those in Fig.~\ref{Fig-Tri-E500-H1-Dist}, the yield can be increased due to photon polarisation. In Fig.~\ref{Fig-Tri-E500-H1-Dist}, because the first Compton harmonic appears at $s_{\gamma,1}\to1$, the effect of the pair suppression due to the photon polarisation in the regime $s_{\gamma}>s_{\gamma,1}$ can be weaker overall than the enhancement due to the photon polarisation in the regime $s_{\gamma}<s_{\gamma,1}$. In contrast, for lower photon energies, as we will show in Fig.~\ref{Fig-Tri-E165-H3-Dist}, the total trident yield can be noticeably reduced by photon polarisation as only photons with $s_{\gamma}>s_{\gamma,1}$ can contribute effectively to the production process.

In Fig.~\ref{Fig-Tri-E165-H3-Dist}, we present the results for a $16.5\,\trm{GeV}$ electron. For this lower energy, harmonic structures in the double differential lightfront spectrum are no longer observable because the main contribution is from higher-order harmonics (around $s_{\gamma}=0.64$, the harmonic order is $n=3$). The contribution from the most visible NLC harmonic at $n=1$, is strongly suppressed. Also in this case, because of the photon polarisation, the yield of pairs is slightly reduced around the central peak in the double-differential spectrum and because the contribution from photons at energies below the first NLC harmonic is so suppressed, the part of the spectrum where photon polarisation \emph{enhances} pair production is also suppressed. As a result, in Fig.~\ref{Fig-Tri-E165-H3-Dist} (d), it can be seen that the integral of the polarised case is now noticeably smaller than the unpolarised case. 
The other notable difference from the high-energy case in Fig.~\ref{Fig-Tri-E80-H3-Dist} is that the photons polarised in the $\parallel$-eigenstate has larger contribution to the total positron yield from the NLC-polarised photons as shown in Fig.~\ref{Fig-Tri-E165-H3-Dist} (d).

\begin{figure}[t!!!!]
 \center{\includegraphics[width=0.45\textwidth]{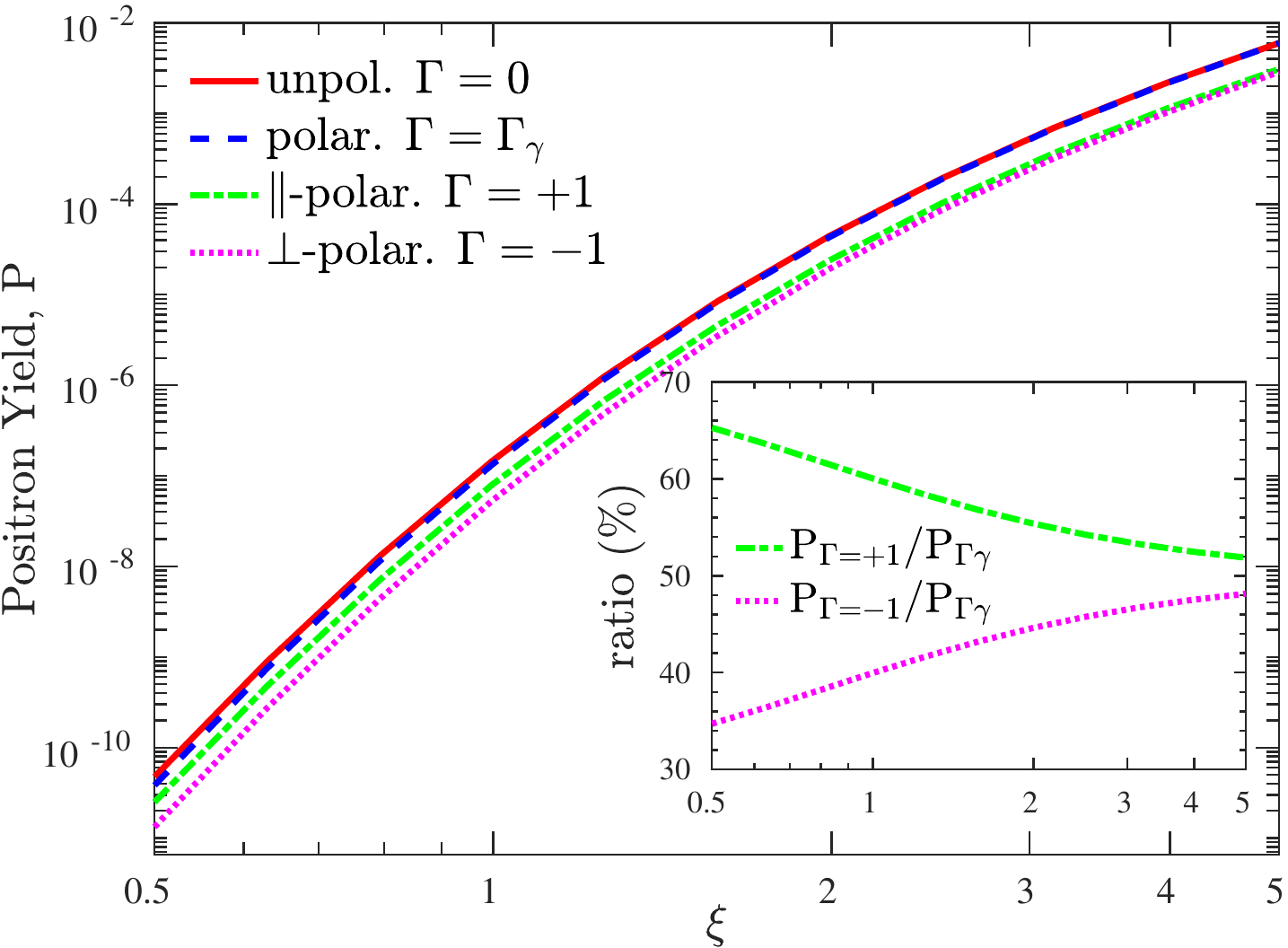}}
\caption{Positron yield as a function of laser intensity for the two-step trident process in a head-on collision of a $16.5\,\trm{GeV}$ electron and a $16$-cycle circularly-polarised plane wave pulse with carrier frequency $\omega_{l}=4.65\,\trm{eV}$ (corresponding to $\eta=0.59$). The plot shows the yield for photons that are unpolarised ($\Gamma=0$), polarised by NLC ($\Gamma=\Gamma_{\gamma}$) or polarised in one of the eigenstates of the laser background ($\Gamma=\pm1$). 
The inset shows the ratio of the yield for photons in a polarisation eigensate to the total yield for the NLC-polarised photons.
}
\label{Fig-Trident-LMA-E165-H3-Numb}
\end{figure}
In~\figref{Fig-Trident-LMA-E165-H3-Numb} the total yield of positrons is calculated as a function of intensity for various photon polarisation choices.
With the increase of the laser intensity, the positron yield increases significantly.
For single-vertex processes, it is known that as $\xi$ increases, the LMA tends to the LCFA for a plane-wave background \cite{Heinzl:2020ynb}.
Since a locally constant field is the same for different circularly-polarised background helicities, we expect that as $\xi$ increases, the difference between each photon helicity state should decrease. 
This is indeed what we observe in the inset in Fig.~\ref{Fig-Trident-LMA-E165-H3-Numb}.
We also observe the converse: as $\xi$ is reduced past $\xi=1$, the relative importance of photon helicity increases. 
This is another example of physics that is beyond the locally constant-field approach.
(The relative importance of the photon polarisation with the change of the laser intensity for the two-step trident process will be discussed later in Fig.~\ref{Fig-Trident-LMA-E165-H3-Numb_Rela_lin_cir}.)

\subsection{Linearly polarised background}
The LMA for a linearly-polarised background is more computationally expensive to calculate than the LMA for a circularly-polarised background due to the latter having azimuthal symmetry in the transverse plane. As a consequence, one must calculate generalised Bessel functions \cite{ritus85,Korsch_2006,PhysRevE.79.026707} in the integrand for a linearly-polarised background, compared to just standard Bessel functions of the first kind for a circularly-polarised background. 
\begin{figure}[t!!!!!]
\includegraphics[width=8cm]{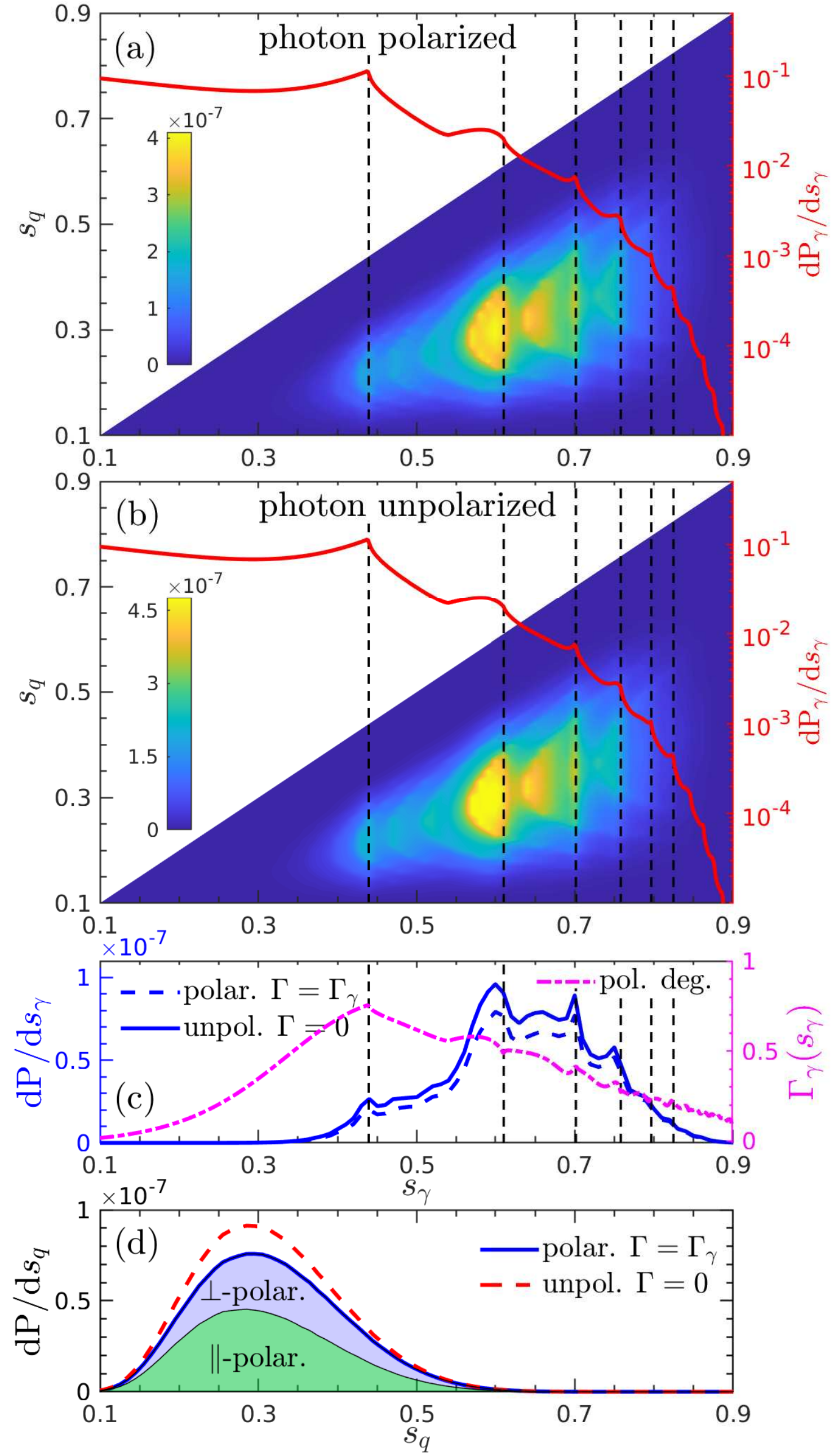}
\caption{Differential probability for the two-step trident process in a $\xi=1$, $N=16$, \emph{linearly} polarised pulse of frequency $4.65\,\trm{eV}$ colliding head on with a $16.5\,\trm{GeV}$ electron, corresponding to an energy parameter of $\eta= 0.59$. The double lightfront spectra, $\ud^{2}\trm{P}/\ud s_{q}/\ud s_{\gamma}$ are plotted for an  (a) polarised and (b) unpolarised photon and compared to the NLC spectrum (red overlaid line). (c) Dependence of the probability on the lightfront momentum of a polarised (blue dashed line) and unpolarised (blue sold line) photon: for the polarised case, the polarisation degree is given by the magenta dashed-dotted line.
(d) Energy spectra of the positron, $\ud\trm{P}/\ud s_{q}$ in the two-step trident process for the unpolarised ($\Gamma=0$) and polarised ($\Gamma=\Gamma_{\gamma}$) photon. The shaded areas underneath the polarised spectrum denote the contributions from each polarisation eigenstate. 
In (a)-(c), the black dashed lines give the location of the first six harmonics in the photon spectrum at $s_{\gamma}=2n\eta /(2n\eta + 1 + \xi^2)$ for integers $n=1$ to $n=6$.
}
\label{Fig-Tri-QED-LMA-E165-H3-Dist-lin}
\end{figure}

Here we perform a similar analysis as in the previous section but focus on the case of a $16.5\,\trm{GeV}$ electron colliding head-on with a $\xi=1$, $N=16$ \emph{linearly} polarised pulse of frequency $4.65\,\trm{eV}$, corresponding to an energy parameter of $\eta\approx 0.59$.
In Fig.~\ref{Fig-Tri-QED-LMA-E165-H3-Dist-lin} rich harmonic structures are observed in the double differential plots, $\ud^{2}\prob/\ud s_{p}/\ud s_{\gamma}$ for a polarised [Fig.~\ref{Fig-Tri-QED-LMA-E165-H3-Dist-lin}\,(a)] and unpolarised [Fig.~\ref{Fig-Tri-QED-LMA-E165-H3-Dist-lin}\,(b)] intermediate photon. 
Again, these harmonic structures correspond to harmonics in the NLC spectrum and cannot be observed in the energy spectrum of the produced pair. Similar to the circularly polarised case, the effect of the photon polarisation decreases the height of the harmonic peak in the double differential plots.
However, in linearly polarised laser backgrounds, the photon is always more likely to be produced in the $\parallel$ polarisation state ($\Gamma_{\gamma}>0$), shown by the magenta line in Fig.~\ref{Fig-Tri-QED-LMA-E165-H3-Dist-lin} (c). Therefore, even with very high-energy electrons, in a linearly-polarised background, the effect of photon polarisation is always to \emph{reduce} the probability of nonlinear trident in the two-step process.
Similar to the low-energy case in circularly polarised backgrounds, the photons in the $\parallel$-polarisation eigenstate has a larger contribution to the total positron yield for the NLC-polarised photons in Fig.~\ref{Fig-Tri-QED-LMA-E165-H3-Dist-lin} (d).

\begin{figure}[t!!!!]
 \center{\includegraphics[width=0.45\textwidth]{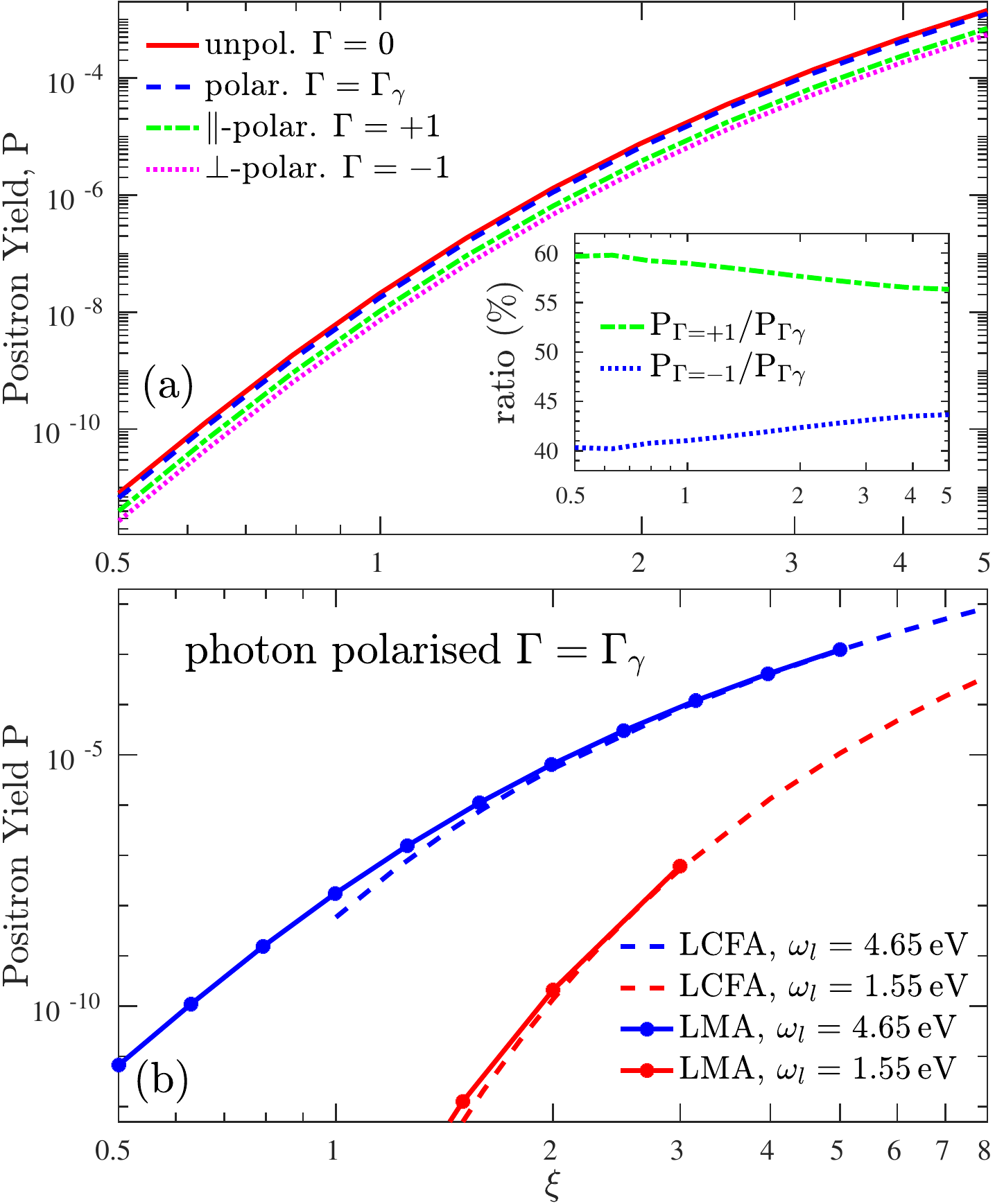}}
\caption{Positron yield as a function of laser intensity for the two-step trident process in a head-on collision of a $16.5\,\trm{GeV}$ electron and a $16$-cycle \emph{linearly}-polarised plane wave pulse with carrier frequency $\omega=4.65\,\trm{eV}$,corresponding to $\eta=0.59$. In plot (a), the yield for photons that are unpolarised ($\Gamma=0$), polarised by NLC ($\Gamma=\Gamma_{\gamma}$) or polarised in one of the eigenstates of the laser background ($\Gamma=\pm1$) is shown; 
the inset shows the ratio of the yield for photons produced in a polarisation eigenstate to the total yield for the NLC-polarised photons. In plot (b), the locally monochromatic approximation is compared to the locally constant field approximation for different laser carrier frequencies.
}
\label{Fig-Trident-LMA-E165-H3-Numb_lin}
\end{figure}
The positron yield is plotted against the background intensity parameter $\xi$ in~\figref{Fig-Trident-LMA-E165-H3-Numb_lin}. 
As in the circularly polarised case, the positron yields for the unpolarised ($\Gamma=0$) and polarised ($\Gamma=\Gamma_{\gamma}$) photon increase significantly with laser intensity, and in the photon-polarised case, the difference between the yield for each photon polarisation eigenstate becomes smaller at higher intensities.
However, in the linearly polarised background, the decrease of this difference is much slower than that in the circularly polarised background, which implies that the polarisation effects in linearly polarised backgrounds persist at much higher laser intensities (we will investigate this in the discussion in Sec. III).
As the direction of polarisation does not change with the phase in linearly polarised backgrounds, there is a well-defined constant field limit. In Fig.~\ref{Fig-Trident-LMA-E165-H3-Numb_lin} (b), the difference between the LMA and LCFA for predicting the positron yield is plotted. 
This illustrates the result that the LMA agrees with the LCFA at large intensity parameter $\xi$
and at small intensities the LMA predicts higher positron yield than the LCFA.
The threshold intensity for the agreement between the LMA and LCFA depends on the frequency of the background field: for the larger laser frequency, the LMA result matches the LCFA at higher intensities.

\section{Discussion}
One of the motivations for the current work was to investigate harmonic structure in the two-step trident process. We found that the harmonic structure from the Compton step is generally washed out in the pair creation step, because the photon energies most likely to create pairs correspond to higher Breit-Wheeler harmonics where harmonic structure is no longer visible. The implications of this for $n$-step higher-order processes and cascades, is that harmonic structure from lower-order processes becomes washed out in later generations due to the lightfront momentum being reduced. For a pair-creation stage, this is clear: to see harmonic structure requires the energy parameter to be of the order of $2(1+\xi^{2})$; already for $\xi=1$ and a triple-harmonic laser frequency of $4.65\,\trm{eV}$ this corresponds to around $130\,\trm{GeV}$ electrons. This washing-out of harmonic structure must also apply to multiple nonlinear Compton scattering of hard photons, where the Compton edge is at energy parameter $2\eta^{2}/(2\eta+1+\xi^{2})$; as the energy parameter $\eta$ reduces in each stage of the cascade, the Compton edge is pushed to lower energies, which are less likely to seed the next generation of the cascade. In order to obtain harmonic structures in later generations, which carry information of the subprocesses in earlier generations, the energy parameter needs to be very high. We demonstrated this for the two-step trident process by considering a $500\,\trm{GeV}$ electron in a circularly-polarised background with frequency  $1.55\,\trm{eV}$. 

Another motivation for the current work was to study the effect due to the intermediate photon being polarised. It is known from past studies of the trident process \cite{king13a,king13b} that in a constant crossed field the relative importance of photon polarisation is an order $10\%$ effect. Indeed this is what we found using the LMA when the intensity parameter was increased towards the $\xi\gg 1$ region. However, with the LMA, we could also study the effect of photon polarisation in the range when $\xi \not \gg 1$, which is the parameter regime of some topical experiments E320 and LUXE that will employ electron beams from a linac.

\begin{figure}[t!!!!]
\centering
\includegraphics[width=8cm]{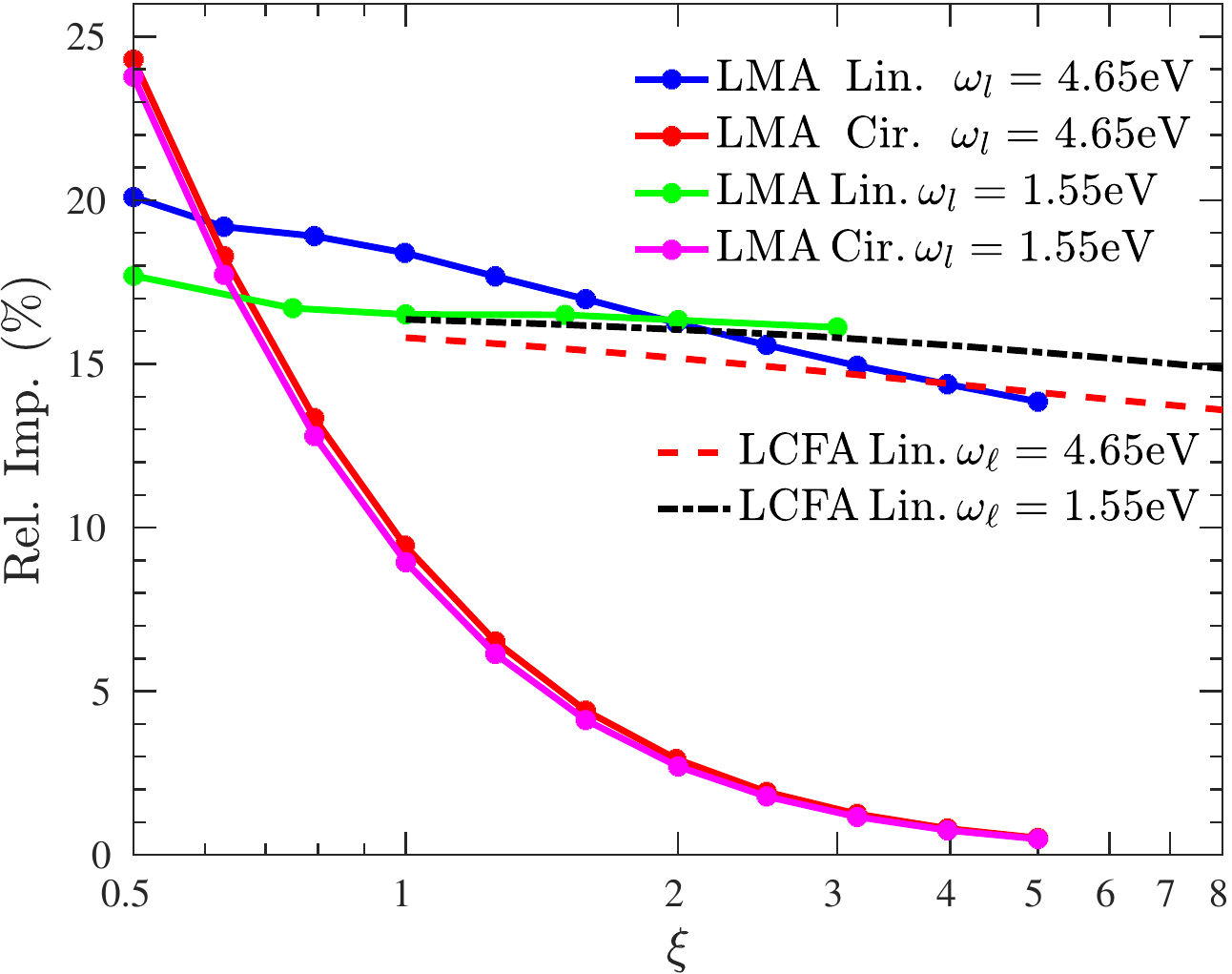}
\caption{The dependency on the intensity parameter of the `relative importance':
$(\prob^{\Gamma=0}-\prob^{\Gamma=\Gamma_{\gamma}})/\prob^{\Gamma=\Gamma_{\gamma}}$
of the polarisation of the intermediate photon in the two-step trident process is illustrated for differently polarised laser backgrounds. $N=16$, $E_{e}=16.5\,\trm{GeV}$.}
\label{Fig-Trident-LMA-E165-H3-Numb_Rela_lin_cir}
\end{figure}
In Fig.~\ref{Fig-Trident-LMA-E165-H3-Numb_Rela_lin_cir}, we compare the relative importance of the polarisation property of the intermediate photon in the two-step trident process in differently polarised laser backgrounds. In a \emph{linearly}-polarised background, the effect of photon polarisation in the two-step trident for likely future experimental parameters, is to \emph{lower} the trident rate by around $15\%$ in the intermediate intensity regime of $\xi \sim O(1)$. This agrees with a similar analysis performed for a constant crossed field background \cite{king13a}. 
In a linearly-polarised background, the direction of polarisation does not change with phase in the lab frame, and therefore there is a well-defined constant field limit. However, for circular polarisation, there is no well-defined constant-crossed field limit. Therefore, if the LCFA becomes more accurate as $\xi$ is increased, the importance of photon polarisation in circular backgrounds must reduce for larger $\xi$, which we indeed find in Fig.~\ref{Fig-Trident-LMA-E165-H3-Numb_Rela_lin_cir}. For $\xi\lesssim 1$ in circular backgrounds, the polarization effect can also reduce the trident rate, in principle even more than $10\%$. 
Here we see the point of using the LMA, which has been used to calculate the polarisation effect in the intermediate intensity region and in field backgrounds with the circular polarisation. Agreement between the LMA and LCFA in the high-intensity region is shown in Fig.~\ref{Fig-Trident-LMA-E165-H3-Numb_Rela_lin_cir}.
Since the applicability of the LCFA also depends on the energy parameter, we include in the plot the leading and third harmonic of the background frequency. We see little difference in the relative importance although the energy parameter differs by a factor of three. This is perhaps due to the fact that, when all other parameters are fixed, for increasing energy parameter, the LCFA for NLC becomes \emph{less} accurate, but the LCFA for NBW becomes \emph{more} accurate.
So whilst linear polarisation is an attractive background to use for trident experiments since, for fixed laser pulse energy, it allows for $\xi$ to be increased by a factor $\sqrt{2}$ compared to a circularly-polarised background, we see that there is a slight cost to this increase when one takes into account the polarisation of intermediate photons.

One may ask the question of whether knowledge of the influence of photon polarisation can be used to optimise the two-step. Since nonlinear Compton most abundantly produces photons in the polarisation state that is least likely to create pairs, if a single laser pulse is used, there is a natural compensation against the effects of polarisation. However, if one uses a double laser pulse, with each sub-pulse being linearly polarised in a plane orthogonal to the polarisation of the other pulse, and if one chooses the properties of the pulses such that in the first pulse mainly only nonlinear Compton scattering takes place, then one can engineer a situation in which the most abundantly produced polarisation in nonlinear Compton scattering is also the polarisation most likely to create pairs in the second pulse. Such an approach can be presumably generalised to $n$-stage processes, depending on which sub-process is favoured in each generation.

\begin{figure}[t!!!]
 \center{\includegraphics[width=0.48\textwidth]{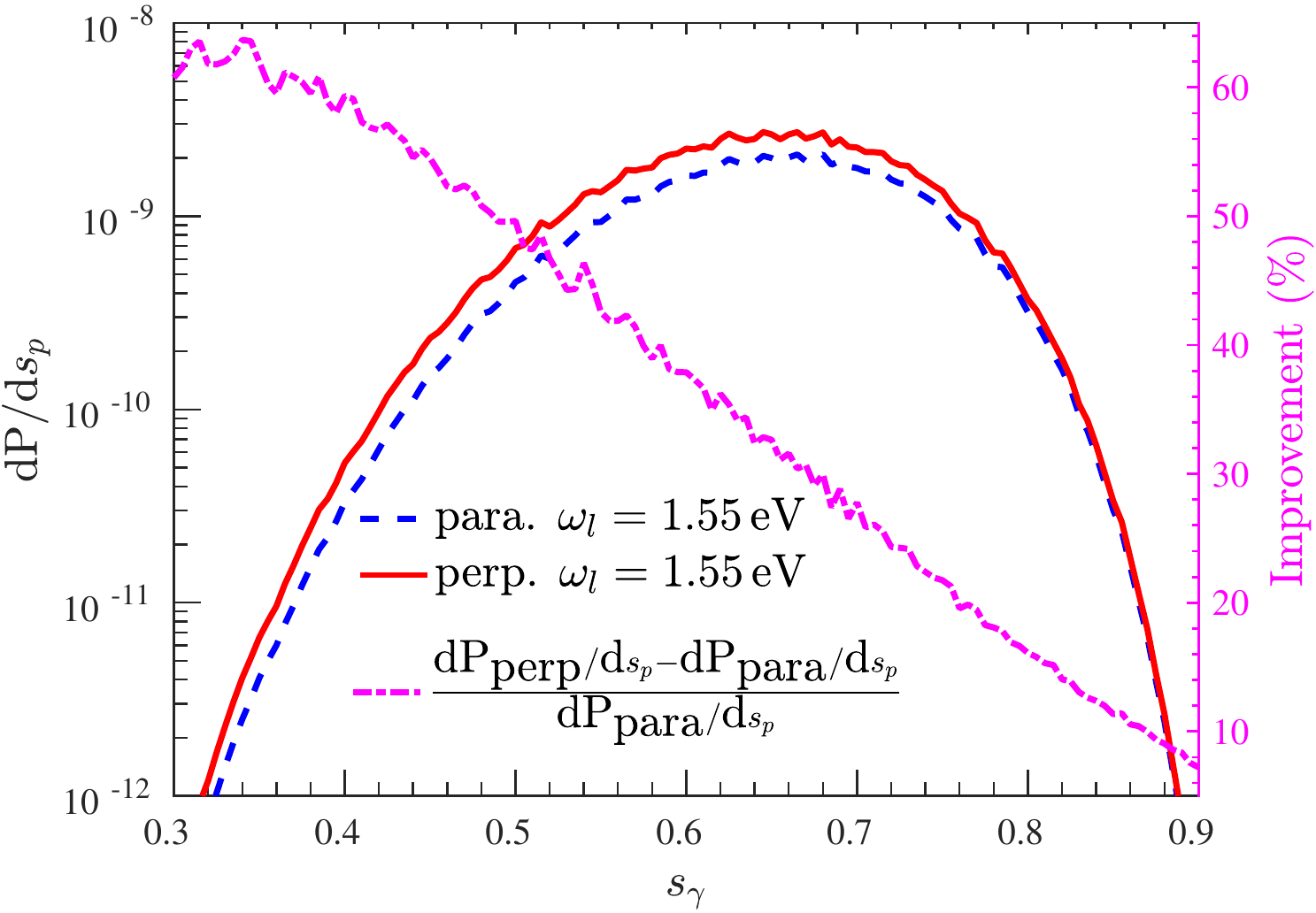}}
\caption{Positron yield in the two-step trident scenario, in which a beam of $16.5\,\trm{GeV}$ electrons collide head-on with one plane-wave pulse to scatter highly polarised $\gamma$ photons via nonlinear Compton scattering, which then collide with a second plane-wave pulse to create pairs via the nonlinear Breit-Wheeler process. Both the plane waves are $16$-cycle linearly polarised pulses with intensity parameter $\xi=2$ and carrier frequency $1.55\,\trm{eV}$, which are polarised in the plane parallel or perpendicular to each other. The relative difference in the positron yield between the two cases is shown by the magenta line. $\trm{P}_{\trm{para}}$ ($\trm{P}_{\trm{perp}}$) denotes the positron yield in the case with the parallel (perpendicular) polarisation.}
\label{Fig-Opt-Spec-NLC-NBW-a2}
\end{figure}

The total yield of pairs in this two-step scenario can be calculated as in the previous sections using \eqnref{eqn:tri2},
but now where the corresponding phase integrals for NLC and NBW correspond to an integral over different laser pulses. In Fig.~\ref{Fig-Opt-Spec-NLC-NBW-a2}, we compare the positron yield in the double laser pulses with the parallel and perpendicular linear polarisation. 
As we can see, by using laser pulses with orthogonal polarisations,
one can improve the positron yield about $32\%$ than that from the two laser pulses with the parallel polarisation.

\section{Conclusion}
In the two-step trident process, photon polarisation can have an O(10\%) effect on the total rate. This conclusion agrees with previous studies in a constant crossed field and via the locally constant field approximation, which one would expect to be a good approximation when $\xi\gg1$. What is new here is that we have used the locally monochromatic approximation, which allowed us to analyse $\xi \sim O(1)$ as well as backgrounds with circularly-polarisation. We find that the effect of photon polarisation increases slightly for a linearly polarised background as $\xi$ is made smaller; for a circularly polarised background, photon polarisation is only $O(10\%)$ for $\xi\lesssim 1$: as $\xi$ is increased above this, the effect of photon polarisation disappears. In order to reach this conclusion we have derived photon-polarised rates for nonlinear Compton scattering and nonlinear Breit-Wheeler pair creation, in both a circularly-polarised and a linearly-polarised background. Such expressions can be useful in extending simulation codes based on the locally monochromatic approximation to include photon-polarised subprocesses.

If the energy of the scattered electrons and positrons were measured in experiment, the double-differential probability can in principle show harmonic structure from the Compton step intersecting with harmonic structure from nonlinear Breit-Wheeler.
However, this can only be achieved if the centre-of-mass energy for a single background photon colliding with the electron is significantly over the threshold for linear Breit-Wheeler. We illustrated this with an example of $500\,\trm{GeV}$ electrons and a $1.55\,\trm{eV}$ laser frequency. In general we conclude that harmonic structure will be washed out in later generations in any $n$-step process such as a QED cascade.

Finally, one way to exploit the dependence on photon polarisation to enhance the two-step trident process, is to collide initial electrons with two linearly-polarised laser pulses, which have perpendicular polarisations. This can lead to an enhancement of around $30\%$ compared to two pulses with the same polarisation.

\section{Acknowledgments}
ST acknowledges support from the Shandong Provincial Natural Science Foundation, Grants No. ZR202102280476. BK
acknowledges the hospitality of the DESY theory group
and support from the Deutsche Forschungsgemeinschaft
(DFG, German Research Foundation) under Germany’s
Excellence Strategy – EXC 2121 “Quantum Universe” –
390833306.
The work was carried out at Marine Big Data Center of Institute for Advanced Ocean Study of Ocean University of China.
\appendix

\begin{widetext}

\section{LMA formulas for polarised nonlinear Compton and Breit-Wheeler}\label{AppA}

All LMA probabilities, $\prob$, can be written as a sum over partial probabilities, $\prob_{n}$, corresponding to integer harmonics $n$, with the lower bound of the sum $\lceil n^{\ast} \rceil$ depending on the process and experimental parameters. The total probability is calculated as an integral over the lightfront spectrum $\ud\tsf{P}/\ud s = \sum_{n=\lceil n^{\ast}(s)\rceil}^{\infty} \ud\tsf{P}_{n}/\ud s$ of one of the outgoing particles.

The photon polarisation eigenstates are defined as:
\begin{align}
\varepsilon^{\mu}_{1,2}=\epsilon^{\mu}_{1,2}-\frac{\ell\cdot \epsilon_{1,2}}{k\cdot \ell}k^{\mu}; \qquad \varepsilon^{\mu}_{\pm}=\epsilon^{\mu}_{\pm}-\frac{\ell\cdot \epsilon_{\pm}}{k\cdot \ell}k^{\mu},
\end{align}
where $\epsilon_{\pm} = (\epsilon_{1} \pm i\epsilon_{2})/\sqrt{2}$ and $\epsilon_{j}^{\mu} = \delta_{j}^{\mu}$. 
where
$\varepsilon^{\mu}_{1,2}$ are the eigenstates in a linearly-polarised plane wave background and $\varepsilon^{\mu}_{\pm}$ those in a circularly-polarised plane-wave background.
The designation: $\parallel$ ($\perp$) refers to $\eps_{1}$ ($\eps_{2}$) in a linearly-polarised background 
and in a circularly-polarised background to photons polarised in the $\eps_{+}$ ($\eps_{-}$) state, which are states with the opposite (same) helicity as the background.

The photon polarisation degree of the photons, can be defined for the entire spectrum, $\Gamma$; for a particular photon energy in the spectrum $\Gamma(s)$; or for a particular photon energy at a given instantaneous value of the phase, $\phi$ as $\Gamma(s,\phi)$. Explicitly, the definitions are:
\bea 
\Gamma = \frac{\prob^{\parallel}-\prob^{\perp}}{\prob^{\parallel}+\prob^{\perp}}; \qquad \Gamma(s) = \frac{\ud\prob^{\parallel}(s)/\ud s-\ud \prob^{\perp}(s)/\ud s}{\ud\prob^{\parallel}(s)/\ud s+\ud \prob^{\perp}(s)/\ud s}; \qquad \Gamma(s,\phi) = \frac{\ud^{2}\prob^{\parallel}(s,\phi)/\ud s\,\ud \phi-\ud^{2}\prob^{\perp}(s,\phi)/\ud s\,\ud \phi}{\ud^{2}\prob^{\parallel}(s,\phi)/\ud s\,\ud \phi+\ud^{2}\prob^{\perp}(s,\phi)/\ud s\,\ud \phi}.
\eea
These relations can be straightforwardly inverted, e.g. $\prob^{\parallel} = \prob(1+\pdeg)/2$,  and $\prob^{\perp} = \prob(1-\pdeg)/2$.

First-order rates are usually defined in terms of the lightfront momentum fraction $s \in [0,1]$ of the incoming particle. For the two-step trident process, in the NBW part, the lightfront momentum fraction of the positron $s_{q}$ is not bounded by $1$, but instead, $s_{q} \in [0,s_{\gamma}]$, as explained in \eqnref{eqn:sdef}. Therefore, to keep these definitions as referring to single-vertex processes, we introduce the lightfront momentum variable for the positron $t=s_{q}/s_{\gamma}$, so that $t\in[0,1]$.

We also recall the definition of the plane-wave potential, which depends upon the phase $\phi=\vkap \cdot x$ in the form:
\bea
\mbf{a}^{\perp} = \begin{cases}
m\xi f(\phi) (\cos \vphi, \varsigma\sin\phi ) & |\phi| < \Phi/2 \nn \\
0 & \trm{otherwise},
\end{cases}
\eea
where $a = (0, \mbf{a}^{\perp},0)$ and $\varsigma \in \{-1,0,1\}$ is chosen to switch between linear ($\varsigma=0$) or circular ($\varsigma=\pm 1$) polarisation.
(In this paper, a circularly polarised background refers to the choice $\varsigma=+1$.)

\subsection{Photon-polarised nonlinear Compton scattering in a circularly-polarised background}
The partial lightfront momentum spectrum, $\ud\tsf{P}^{\LCpara(\LCperp)}_{\cpo,\gamma,n}/\ud s$, of photons in the $\parallel(\perp)$-state emitted by an unpolarised electron with energy parameter $\eta$ via nonlinear Compton scattering in a circularly-polarised plane wave background is given by:
\begin{subequations}
\begin{align}
\frac{\ud \tsf{P}^{\LCpara}_{\cpo,\gamma,n}}{\ud s}=&\frac{\alpha}{2\eta} \int \ud\phi \,\Theta[n-n^{\ast}_{\cpo,\gamma}(\phi)]
\left\{\xi^2(\phi)\left[J'^{2}_{n}(z_{\gamma,n}) + \frac{n^{2}}{(z_{\gamma,n})^{2}}J^{2}_{n}(z_{\gamma,n}) -  J^{2}_n(z_{\gamma,n})\right]h_{s} - J^{2}_{n}(z_{\gamma,n}) \right.\nonumber\\
               &~~~~~~~~~~~~~~~~~~~~~~~~~~~~~~~~~~~\left. + 2 h_{s} \frac{\xi(\phi)}{\mathcal{P}_{\cpo,\gamma,n}}\left(1 + \xi^{2}(\phi) - n\eta\frac{1-s}{s} \right)J'_{n}(z_{\gamma,n})J_{n}(z_{\gamma,n}) \right\}\,,\\
\frac{\ud \tsf{P}^{\LCperp}_{\cpo,\gamma,n}}{\ud s}=&\frac{\alpha}{2\eta} \int \ud\phi \,\Theta[n-n^{\ast}_{\cpo,\gamma}(\phi)]
\left\{\xi^2(\phi)\left[J'^{2}_{n}(z_{\gamma,n}) + \frac{n^{2}}{(z_{\gamma,n})^{2}}J^{2}_{n}(z_{\gamma,n}) -  J^{2}_n(z_{\gamma,n})\right]h_{s} - J^{2}_{n}(z_{\gamma,n}) \right.\nonumber\\
               &~~~~~~~~~~~~~~~~~~~~~~~~~~~~~~~~~~~\left. - 2 h_{s} \frac{\xi(\phi)}{\mathcal{P}_{\cpo,\gamma,n}}\left(1 + \xi^{2}(\phi) - n\eta\frac{1-s}{s} \right)J'_{n}(z_{\gamma,n})J_{n}(z_{\gamma,n}) \right\}\,,
\label{Eq-LMA-prob-s}
\end{align}
\end{subequations}
where $\xi(\phi)\equiv\xi f(\phi)$, and
\[z_{\gamma,n} =  \frac{s \xi(\phi)}{\eta (1-s)}\mathcal{P}_{\cpo,\gamma,n}, \quad
  \mathcal{P}_{\cpo,\gamma,n}  = \sqrt{2n\eta\frac{1-s}{s} - 1 - \xi^{2}(\phi)},\quad
  n^{\cpo}_{\ast,\gamma}(\phi) = \frac{s[1+\xi^{2}(\phi)]}{2\eta(1-s)},\quad
h_{s}=\frac{1+(1-s)^{2}}{2(1-s)}.\]
The total spectrum of the emitted photon from an unpolarised electron in a circularly-polarised background can then be given as
\[\frac{\ud}{\ud s}\tsf{P}_{\cpo,\gamma,n}=\frac{\ud}{\ud s}\tsf{P}^{\LCpara}_{\cpo,\gamma,n}+\frac{\ud}{\ud s}\tsf{P}^{\LCperp}_{\cpo,\gamma,n}.\]

\subsection{Photon-polarised nonlinear Breit-Wheeler pair-creation in a circularly-polarised background}
The partial lightfront momentum spectrum, $\ud\tsf{P}^{\LCpara(\LCperp)}_{\cpo,e,n}/\ud t$ of positrons created via the nonlinear Breit-Wheeler process in a circularly-polarised plane wave background by a photon with energy parameter $\eta_{\ell}$ and polarised in the $\parallel(\perp)$-eigenstate is given by:
\begin{subequations}
\begin{align}
\frac{\ud\tsf{P}^{\LCpara}_{\cpo,e,n}}{\ud t} =&\frac{\alpha}{ \eta_{\ell}}  \int \ud\vphi\,\Theta[n-n^{\ast}_{\cpo,e}(\phi)] \left\{J^{2}_{n}(z_{e,n}) + \xi^{2}(\phi)\left[\frac{n^2}{(z_{e,n})^2} J^{2}_{n}(z_{e,n}) + J'^{2}_{n}(z_{e,n})-J^{2}_{n}(z_{e,n})\right]g_{t} \right.\nonumber\\
&~~~~~~~~~~~~~~~~~~~~~~~~~~~~~~~~~~~~~~\left. + \frac{2\xi(\phi)}{\mathcal{P}_{\cpo,e,n}}\left[n \eta_{\ell} t(1-t) -1 - \xi^{2}(\phi)  \right] J'_{n}(z_{e,n}) J_{n}(z_{e,n})g_{t}\right\}\,,\\
\frac{\ud\tsf{P}^{\LCperp}_{\cpo,e,n}}{\ud t} =&\frac{\alpha}{ \eta_{\ell}}  \int \ud\vphi\,\Theta[n-n^{\ast}_{\cpo,e}(\phi)] \left\{J^{2}_{n}(z_{e,n}) + \xi^{2}(\phi)\left[\frac{n^2}{(z_{e,n})^2} J^{2}_{n}(z_{e,n}) + J'^{2}_{n}(z_{e,n})-J^{2}_{n}(z_{e,n})\right]g_{t} \right.\nonumber\\
&~~~~~~~~~~~~~~~~~~~~~~~~~~~~~~~~~~~~~~\left. - \frac{2\xi(\phi)}{\mathcal{P}_{\cpo,e,n}}\left[n \eta_{\ell} t(1-t) -1 - \xi^{2}(\phi)  \right] J'_{n}(z_{e,n}) J_{n}(z_{e,n})g_{t}\right\}\,,
\label{Eq-NBW-polar-LMA-S}
\end{align}
\end{subequations}
where
\[z_{e,n}=\frac{\xi(\phi)\mathcal{P}_{\cpo,e,n}}{\eta_{\ell} t (1-t)},\quad
 \mathcal{P}_{\cpo,e,n} =\sqrt{2n \eta_{\ell} t (1-t) -1 - \xi^{2}(\phi)},\quad
 n^{\ast}_{\cpo,e}(\phi) = \frac{1 + \xi^{2}(\phi)}{2\eta_{\ell} t (1-t)},\quad
 g_{t} = \frac{t^{2} + (1-t)^2}{2t(1-t)}\,.\]
We can then obtain the positron lightfront spectrum created by an unpolarized photon in a circularly-polarised background as
\[\frac{\ud}{\ud t}\tsf{P}_{\cpo,e,n}=\frac{1}{2}\left(\frac{\ud}{\ud t}\tsf{P}^{\LCpara}_{\cpo,e,n}+\frac{\ud}{\ud t}\tsf{P}^{\LCperp}_{\cpo,e,n}\right).\]

\subsection{Photon-polarised nonlinear Compton scattering in a linearly-polarised background}
The partial lightfront spectrum, $\ud\tsf{P}^{\LCpara(\LCperp)}_{\lpo,\gamma,n}/\ud s$, of photons in the $\parallel(\perp)$-state emitted by an unpolarised electron with energy parameter $\eta$ via nonlinear Compton scattering in a linearly-polarised plane wave background is given by:
\begin{subequations}
\begin{align}
\frac{\ud \tsf{P}^{\LCpara}_{\lpo,\gamma,n}}{\ud s}=&
\frac{\alpha }{2\eta} \int \ud\phi\,\Theta[n-n^{\ast}_{\lpo,\gamma}(\phi)] \int^{\pi}_{-\pi}\frac{\ud \theta}{2\pi} \left\{\xi^{2}(\phi)  \left[\Lambda^{2}_{1,n}(\alpha_{\gamma},\beta_{\gamma}) - \Lambda_{0,n}(\alpha_{\gamma},\beta_{\gamma})\Lambda_{2,n}(\alpha_{\gamma},\beta_{\gamma})\right]h_s- \Lambda^{2}_{0,n}(\alpha_{\gamma},\beta_{\gamma}) \right.\nonumber\\
&~~~~~~~~~~~~~~~~~~~~~~+\left. \left[\Lambda_{0,n}(\alpha_{\gamma},\beta_{\gamma})\mathcal{P}_{\lpo,\gamma,n}\cos \theta-\xi(\phi)\Lambda_{1,n}(\alpha_{\gamma},\beta_{\gamma})\right]^{2}- [\Lambda_{0,n}(\alpha_{\gamma},\beta_{\gamma})\mathcal{P}_{\lpo,\gamma,n}\sin \theta]^{2} \right\}\,,\\
\frac{\ud \tsf{P}^{\LCperp}_{\lpo,\gamma,n}}{\ud s}=&
\frac{\alpha }{2\eta} \int \ud\phi\,\Theta[n-n^{\ast}_{\lpo,\gamma}(\phi)] \int^{\pi}_{-\pi}\frac{\ud \theta}{2\pi} \left\{\xi^{2}(\phi)  \left[\Lambda^{2}_{1,n}(\alpha_{\gamma},\beta_{\gamma}) - \Lambda_{0,n}(\alpha_{\gamma},\beta_{\gamma})\Lambda_{2,n}(\alpha_{\gamma},\beta_{\gamma})\right]h_s- \Lambda^{2}_{0,n}(\alpha_{\gamma},\beta_{\gamma}) \right.\nonumber\\
&~~~~~~~~~~~~~~~~~~~~~~-\left. \left[\Lambda_{0,n}(\alpha_{\gamma},\beta_{\gamma})\mathcal{P}_{\lpo,\gamma,n}\cos \theta-\xi(\phi)\Lambda_{1,n}(\alpha_{\gamma},\beta_{\gamma})\right]^{2}+ [\Lambda_{0,n}(\alpha_{\gamma},\beta_{\gamma})\mathcal{P}_{\lpo,\gamma,n}\sin \theta]^{2} \right\}\,,
\end{align}
\label{Eq-NLC-LMA-lin}
\end{subequations}
where 
\[ \mathcal{P}_{\lpo,\gamma,n}   = \sqrt{2n\eta \frac{1-s}{s} - 1 - \frac{1}{2}\xi^{2}(\phi)},\quad
n^{\ast}_{\lpo,\gamma}(\phi) = \frac{s[1+\xi^{2}(\phi)/2]}{2\eta(1-s)},\quad
\alpha_{\gamma}= \mathcal{P}_{\lpo,\gamma,n}\frac{s\xi(\phi)\cos \theta}{\eta (1-s)},\quad
\beta_{\gamma}= \frac{s\,\xi^{2}(\phi)}{8\eta (1-s)}\,.\]
and
\begin{subequations}
\begin{align}
\Lambda_{0,n}(x,y)&=\sum_{k=-\infty}^{\infty}J_{n+2k}(x)J_{k}(y)\,,\\
\Lambda_{1,n}(x,y)&=\frac{1}{x}\sum_{k=-\infty}^{\infty}(n+2k)J_{n+2k}(x)J_{k}(y)\\
\Lambda_{2,n}(x,y)&=\frac{1}{2}\sum_{k=-\infty}^{\infty}J_{n+2k}(x)\left(\frac{k}{y}+1\right)J_{k}(y)\,,
\end{align}
\label{Eq-LMA-lin-harm}
\end{subequations}
The total spectrum of the emitted photon from an unpolarised electron can then be given as
\[\frac{\ud \tsf{P}_{\lpo,\gamma,n}}{\ud s}=\frac{\ud \tsf{P}^{\LCpara}_{\lpo,\gamma,n}}{\ud s}+\frac{\ud \tsf{P}^{\LCperp}_{\lpo,\gamma,n}}{\ud s}.\]

\subsection{Photon-polarised nonlinear Breit-Wheeler pair-creation in a linearly-polarised background}
The partial lightfront momentum spectrum, $\ud\tsf{P}^{\LCpara(\LCperp)}_{\lpo,e,n}/\ud t$ of positrons created via the nonlinear Breit-Wheeler process in a linearly-polarised plane wave background by a photon with energy parameter $\eta_{\ell}$ and polarised in the $\parallel(\perp)$-state is given by:
\begin{subequations}
\begin{align}
\frac{\ud\tsf{P}^{\LCpara}_{\lpo,e,n}}{\ud t}&=\frac{\alpha}{\eta_{\ell}} \int \ud\phi\,\Theta[n-n^{\ast}_{\lpo,e}(\phi)]  \int^{\pi}_{-\pi}\frac{\ud \theta}{2\pi}\left\{\xi^{2}(\phi)\left[\Lambda^{2}_{1,n}(\alpha_{e},\beta_{e}) -\Lambda_{0,n}(\alpha_{e},\beta_{e})\Lambda_{2,n}(\alpha_{e},\beta_{e})\right]g_{t} + \Lambda^{2}_{0,n}(\alpha_{e},\beta_{e})\right.\nonumber\\
       &\left.~~~~~~~~~~~~~~~~~~~~~~~~~~ + \left[\Lambda_{0,n}(\alpha_{e},\beta_{e})\mathcal{P}_{\lpo,e,n} \sin\theta\right]^{2} - \left[\Lambda_{0,n}(\alpha_{e},\beta_{e})\mathcal{P}_{\lpo,e,n} \cos\theta - \xi(\phi) \Lambda_{1,n}(\alpha_{e},\beta_{e})\right]^{2} \right\}\,,\\
\frac{\ud\tsf{P}^{\LCperp}_{\lpo,e,n}}{\ud t}&=\frac{\alpha}{\eta_{\ell}} \int \ud\phi\,\Theta[n-n^{\ast}_{\lpo,e}(\phi)] \int^{\pi}_{-\pi}\frac{\ud \theta}{2\pi}\left\{\xi^{2}(\phi)\left[\Lambda^{2}_{1,n}(\alpha_{e},\beta_{e}) -\Lambda_{0,n}(\alpha_{e},\beta_{e})\Lambda_{2,n}(\alpha_{e},\beta_{e})\right]g_{t} + \Lambda^{2}_{0,n}(\alpha_{e},\beta_{e})\right.\nonumber\\
       &\left.~~~~~~~~~~~~~~~~~~~~~~~~~~ - \left[\Lambda_{0,n}(\alpha_{e},\beta_{e})\mathcal{P}_{\lpo,e,n} \sin\theta\right]^{2} + \left[\Lambda_{0,n}(\alpha_{e},\beta_{e})\mathcal{P}_{\lpo,e,n} \cos\theta - \xi(\phi) \Lambda_{1,n}(\alpha_{e},\beta_{e})\right]^{2} \right\}\,,
\label{Eq_NBW_LMA_lin}
\end{align}
\end{subequations}
where the expressions of $\Lambda_{j,n}$ for $j\in\{0,1,2\}$ are the same as in Eq.~(\ref{Eq-LMA-lin-harm}), and
\[\mathcal{P}_{\lpo,e,n} = \sqrt{2n\eta_{\ell} t (1-t) - 1 - \frac{1}{2}\xi^{2}(\phi)},\quad
n^{\ast}_{\lpo,e}(\phi)=\frac{1+\xi^{2}(\phi)/2}{2\eta_{\ell} t(1-t)},\quad
\alpha_{e}= \mathcal{P}_{\lpo,e,n}\frac{\xi(\phi)  \cos\theta}{\eta_{\ell} (1-t)t},\quad
\beta_{e}=\frac{\xi^{2}(\phi)}{8 t\eta_{\ell} (1-t)}.\]
The partial lightfront momentum spectrum of the positrons created by an unpolarized photon in a linearly-polarised plane wave background is then given by
\[\frac{\ud\tsf{P}_{\lpo,e,n}}{\ud t}=\frac{1}{2}\left(\frac{\ud\tsf{P}^{\LCpara}_{\lpo,e,n}}{\ud t}+\frac{\ud\tsf{P}^{\LCperp}_{\lpo,e,n}}{\ud t}\right).\]
\end{widetext}

\section{Bandwidth effect at lower electron energies} \label{app:lowEnergy}
As mentioned in the introduction, when the initial electron energy parameter $\eta$ is sufficiently low, bandwidth effects become important in the process of pair creation, and therefore also in the trident process. If the strong-field parameter $\chi=\eta\xi$ satisfies $\chi \ll 1$, pair-creation is strongly suppressed. This suppression can be partially reduced due to the contribution from linear Breit-Wheeler using photons from the upper half of the bandwidth of the pulse. Essentially, there are two routes to pair-creation: i) via nonlinear Breit-Wheeler with the background photons with energies at the carrier frequency; ii) via linear Breit-Wheeler with high energy background photons, which are in the suppressed upper wings of the pulse bandwidth. For $\chi\sim O(1)$, nonlinear Breit-Wheeler is dominant for the parameters studied in this paper. As $\chi$ is reduced and enters the $\chi<1$ region, eventually, nonlinear Breit-Wheeler is suppressed as strongly or even more strongly than the linear Breit-Wheeler contribution, which can become dominant.
This was shown in e.g. \cite{PhysRevLett.108.240406,seipt12b}, and is an artefact due to the theoretical description of the pulse (in reality, such high energy photons from the pulse would not be transmitted through all the optical elements of an experiment). This bandwidth effect was also recently studied in relation to the LMA for Breit-Wheeler \cite{Blackburn:2021cuq,Tang:2021qht} and also nonlinear Compton scattering \cite{King:2020hsk,Blackburn:2021rqm}.

\begin{figure}[h]
         \includegraphics[width=8cm]{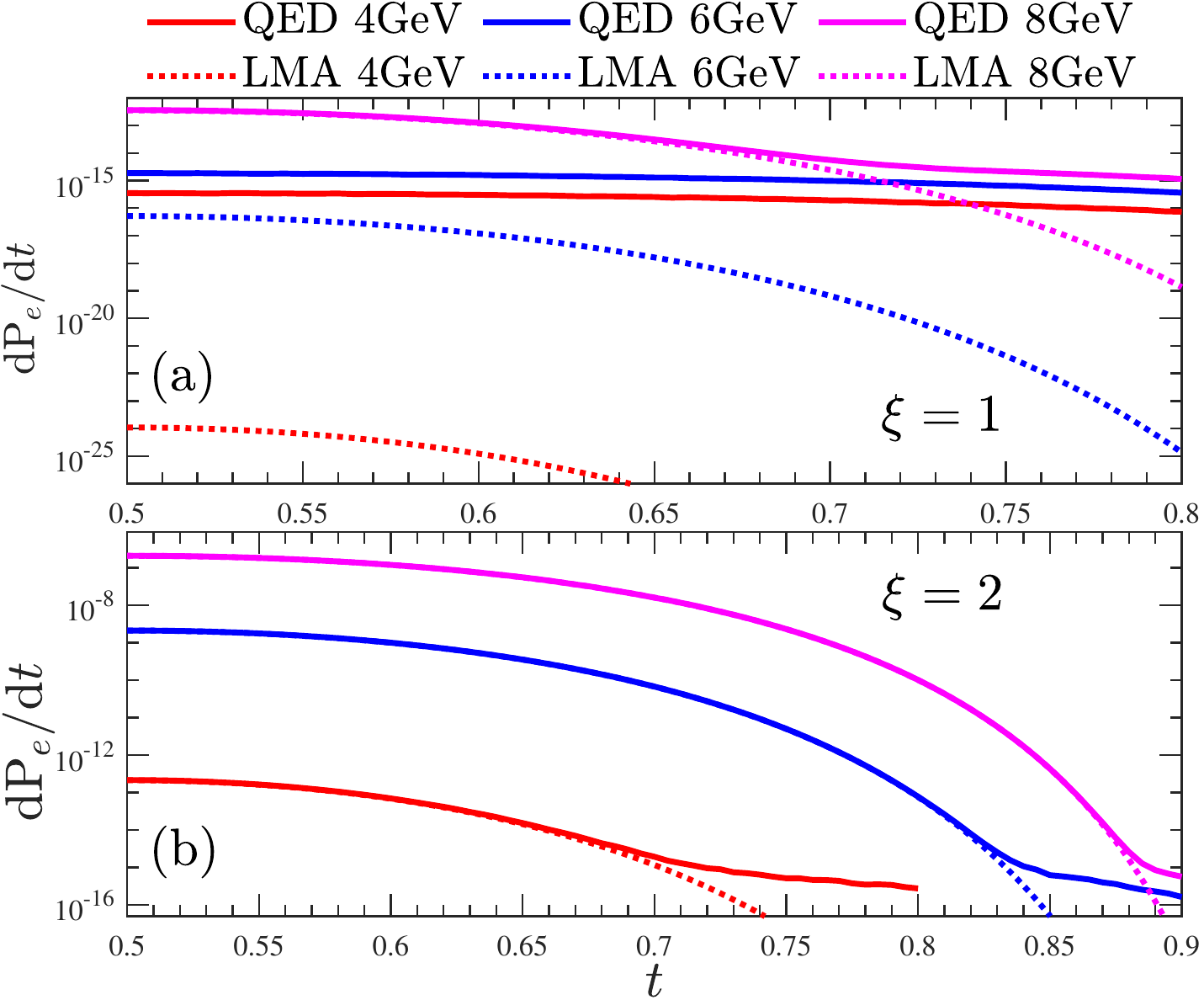}
\caption{Benchmark of the LMA positron spectra $\ud\prob_{e}/\ud t$
with the full QED calculations for the nonlinear Breit-Wheeler pair production in a circularly polarised laser background for different photon energies. The laser pulse has the carrier frequency $\omega_{l}=1.55\,\trm{eV}$, $16$ cycles and an amplitude of (a)  $\xi=1$; and (b) $\xi=2$. The fraction of the lightfront momentum taken by the positron from the incident photon is denoted in this section only, by $s_{q}$.
}
\label{fig:lowEnergy}
\end{figure}

In relation to the trident process, the finite bandwidth of the pulse mostly affects the nonlinear Breit-Wheeler step. 
In \figref{fig:lowEnergy}, we compare the positron lightfront momentum spectra $\ud\prob_{e}/\ud t$ acquired from the full QED expression and the LMA calculations, where we recall that $t$ is the fraction of the lightfront momentum taken by the positron from the incident photon.
We see that, as the photon energy is reduced, the LMA becomes increasingly inaccurate. 
To interpret the accuracy for predicting the two-step trident process, one needs to take into account that the photon energy emitted in the first step that is most likely to create pairs, is at around half  of the electron energy (for the parameters studied in this paper, it is at around 60\% of the electron energy). 
Therefore if $\xi=1$ and the LMA starts to become inaccurate for, say $8\,\trm{GeV}$ photons for the NBW process, this means that the LMA for the two-step trident will likely become inaccurate for
an initial electron energy lower than $16\,\trm{GeV}$ (for the specific parameters used here, we estimate this to be around $13.3\,\trm{GeV}$). This cutoff energy, below which benchmarking the LMA with the full QED expression loses its applicability, depends on $\chi$; if $\xi$ and $\eta$ (or $\omega_{l}$) were increased, the cutoff energy would be lower as can be seen from  Fig.~\ref{fig:lowEnergy}.

We emphasise that this discrepancy between the full QED and the LMA approach is a theoretical artefact and would not be an issue in a real experiment. Even if an experiment were deliberately made to transmit the high-frequency components of the laser pulse, the region where the LMA becomes inaccurate corresponds to very few pairs being created, and so is practically experimentally inaccessible.

\bibliography{current}

\providecommand{\noopsort}[1]{}
\begin{thebibliography}{69}
\expandafter\ifx\csname natexlab\endcsname\relax\def\natexlab#1{#1}\fi
\expandafter\ifx\csname bibnamefont\endcsname\relax
  \def\bibnamefont#1{#1}\fi
\expandafter\ifx\csname bibfnamefont\endcsname\relax
  \def\bibfnamefont#1{#1}\fi
\expandafter\ifx\csname citenamefont\endcsname\relax
  \def\citenamefont#1{#1}\fi
\expandafter\ifx\csname url\endcsname\relax
  \def\url#1{\texttt{#1}}\fi
\expandafter\ifx\csname urlprefix\endcsname\relax\def\urlprefix{URL }\fi
\providecommand{\bibinfo}[2]{#2}
\providecommand{\eprint}[2][]{\url{#2}}

\bibitem[{\citenamefont{Bula et~al.}(1996)}]{E144:1996enr}
\bibinfo{author}{\bibfnamefont{C.}~\bibnamefont{Bula}} \bibnamefont{et~al.}
  (\bibinfo{collaboration}{E144}), \bibinfo{journal}{Phys. Rev. Lett.}
  \textbf{\bibinfo{volume}{76}}, \bibinfo{pages}{3116} (\bibinfo{year}{1996}).

\bibitem[{\citenamefont{Burke et~al.}(1997)}]{burke97}
\bibinfo{author}{\bibfnamefont{D.~L.} \bibnamefont{Burke}}
  \bibnamefont{et~al.}, \bibinfo{journal}{Phys. Rev. Lett.}
  \textbf{\bibinfo{volume}{79}}, \bibinfo{pages}{1626} (\bibinfo{year}{1997}).

\bibitem[{\citenamefont{Bamber et~al.}(1999)}]{bamber99}
\bibinfo{author}{\bibfnamefont{C.}~\bibnamefont{Bamber}} \bibnamefont{et~al.},
  \bibinfo{journal}{Phys. Rev. D} \textbf{\bibinfo{volume}{60}},
  \bibinfo{pages}{092004} (\bibinfo{year}{1999}).

\bibitem[{\citenamefont{Nielsen et~al.}(2022)\citenamefont{Nielsen, Holtzapple,
  Lund, Surrow, S\o{}rensen, and Uggerh\o{}j}}]{Nielsen:2022bws}
\bibinfo{author}{\bibfnamefont{C.~F.} \bibnamefont{Nielsen}},
  \bibinfo{author}{\bibfnamefont{R.}~\bibnamefont{Holtzapple}},
  \bibinfo{author}{\bibfnamefont{M.~M.} \bibnamefont{Lund}},
  \bibinfo{author}{\bibfnamefont{J.~H.} \bibnamefont{Surrow}},
  \bibinfo{author}{\bibfnamefont{A.~H.} \bibnamefont{S\o{}rensen}},
  \bibnamefont{and} \bibinfo{author}{\bibfnamefont{U.~I.}
  \bibnamefont{Uggerh\o{}j}} (\bibinfo{year}{2022}), \eprint{2211.02390}.

\bibitem[{\citenamefont{Chen et~al.}(2022)\citenamefont{Chen, Meuren,
  Gerstmayr, Yakimenko, Bucksbaum, and and}}]{chen22}
\bibinfo{author}{\bibfnamefont{Z.}~\bibnamefont{Chen}},
  \bibinfo{author}{\bibfnamefont{S.}~\bibnamefont{Meuren}},
  \bibinfo{author}{\bibfnamefont{E.}~\bibnamefont{Gerstmayr}},
  \bibinfo{author}{\bibfnamefont{V.}~\bibnamefont{Yakimenko}},
  \bibinfo{author}{\bibfnamefont{P.~H.} \bibnamefont{Bucksbaum}},
  \bibnamefont{and} \bibinfo{author}{\bibfnamefont{D.~A.~R.}
  \bibnamefont{and}}, in \emph{\bibinfo{booktitle}{Optica High-brightness
  Sources and Light-driven Interactions Congress 2022}}
  (\bibinfo{publisher}{Optica Publishing Group}, \bibinfo{year}{2022}), p.
  \bibinfo{pages}{HF4B.6},
  \urlprefix\url{http://opg.optica.org/abstract.cfm?URI=HILAS-2022-HF4B.6}.

\bibitem[{\citenamefont{Abramowicz et~al.}(2021)}]{Abramowicz:2021zja}
\bibinfo{author}{\bibfnamefont{H.}~\bibnamefont{Abramowicz}}
  \bibnamefont{et~al.}, \bibinfo{journal}{Eur. Phys. J. ST}
  \textbf{\bibinfo{volume}{230}}, \bibinfo{pages}{2445} (\bibinfo{year}{2021}),
  \eprint{2102.02032}.

\bibitem[{\citenamefont{Hu et~al.}(2010)\citenamefont{Hu, M{\"u}ller, and
  Keitel}}]{hu10}
\bibinfo{author}{\bibfnamefont{H.}~\bibnamefont{Hu}},
  \bibinfo{author}{\bibfnamefont{C.}~\bibnamefont{M{\"u}ller}},
  \bibnamefont{and} \bibinfo{author}{\bibfnamefont{C.~H.}
  \bibnamefont{Keitel}}, \bibinfo{journal}{Phys. Rev. Lett.}
  \textbf{\bibinfo{volume}{105}}, \bibinfo{pages}{080401}
  (\bibinfo{year}{2010}).

\bibitem[{\citenamefont{Ilderton}(2011)}]{ilderton11}
\bibinfo{author}{\bibfnamefont{A.}~\bibnamefont{Ilderton}},
  \bibinfo{journal}{Phys. Rev. Lett.} \textbf{\bibinfo{volume}{106}},
  \bibinfo{pages}{020404} (\bibinfo{year}{2011}).

\bibitem[{\citenamefont{Hu and Huang}(2014)}]{Hu:2013yrz}
\bibinfo{author}{\bibfnamefont{H.}~\bibnamefont{Hu}} \bibnamefont{and}
  \bibinfo{author}{\bibfnamefont{J.}~\bibnamefont{Huang}},
  \bibinfo{journal}{Phys. Rev. A} \textbf{\bibinfo{volume}{89}},
  \bibinfo{pages}{033411} (\bibinfo{year}{2014}), \eprint{1308.5324}.

\bibitem[{\citenamefont{Dinu and Torgrimsson}(2018)}]{Dinu:2017uoj}
\bibinfo{author}{\bibfnamefont{V.}~\bibnamefont{Dinu}} \bibnamefont{and}
  \bibinfo{author}{\bibfnamefont{G.}~\bibnamefont{Torgrimsson}},
  \bibinfo{journal}{Phys. Rev. D} \textbf{\bibinfo{volume}{97}},
  \bibinfo{pages}{036021} (\bibinfo{year}{2018}), \eprint{1711.04344}.

\bibitem[{\citenamefont{Mackenroth and Di~Piazza}(2018)}]{Mackenroth:2018smh}
\bibinfo{author}{\bibfnamefont{F.}~\bibnamefont{Mackenroth}} \bibnamefont{and}
  \bibinfo{author}{\bibfnamefont{A.}~\bibnamefont{Di~Piazza}},
  \bibinfo{journal}{Phys. Rev. D} \textbf{\bibinfo{volume}{98}},
  \bibinfo{pages}{116002} (\bibinfo{year}{2018}), \eprint{1805.01731}.

\bibitem[{\citenamefont{Dinu and Torgrimsson}(2020)}]{Dinu:2019wdw}
\bibinfo{author}{\bibfnamefont{V.}~\bibnamefont{Dinu}} \bibnamefont{and}
  \bibinfo{author}{\bibfnamefont{G.}~\bibnamefont{Torgrimsson}},
  \bibinfo{journal}{Phys. Rev. D} \textbf{\bibinfo{volume}{101}},
  \bibinfo{pages}{056017} (\bibinfo{year}{2020}), \eprint{1912.11017}.

\bibitem[{\citenamefont{Torgrimsson}(2022)}]{Torgrimsson:2022ndq}
\bibinfo{author}{\bibfnamefont{G.}~\bibnamefont{Torgrimsson}}
  (\bibinfo{year}{2022}), \eprint{2207.05031}.

\bibitem[{\citenamefont{Baier et~al.}(1972)\citenamefont{Baier, Katkov, and
  Strakhovenko}}]{baier72}
\bibinfo{author}{\bibfnamefont{V.~N.} \bibnamefont{Baier}},
  \bibinfo{author}{\bibfnamefont{V.~M.} \bibnamefont{Katkov}},
  \bibnamefont{and} \bibinfo{author}{\bibfnamefont{V.~M.}
  \bibnamefont{Strakhovenko}}, \bibinfo{journal}{Soviet Phys. Nucl. Phys}
  \textbf{\bibinfo{volume}{14}}, \bibinfo{pages}{572} (\bibinfo{year}{1972}).

\bibitem[{\citenamefont{Ritus}(1972)}]{ritus72}
\bibinfo{author}{\bibfnamefont{V.~I.} \bibnamefont{Ritus}},
  \bibinfo{journal}{Nucl. Phys. B} \textbf{\bibinfo{volume}{44}},
  \bibinfo{pages}{236} (\bibinfo{year}{1972}).

\bibitem[{\citenamefont{Morozov and Narozhnyi}(1977)}]{morozov77}
\bibinfo{author}{\bibfnamefont{D.~A.} \bibnamefont{Morozov}} \bibnamefont{and}
  \bibinfo{author}{\bibfnamefont{N.~B.} \bibnamefont{Narozhnyi}},
  \bibinfo{journal}{Zh. Eksp. Teor. Fiz.} \textbf{\bibinfo{volume}{30}},
  \bibinfo{pages}{44} (\bibinfo{year}{1977}),
  \urlprefix\url{http://inspirehep.net/record/123701}.

\bibitem[{\citenamefont{King and Ruhl}(2013)}]{king13b}
\bibinfo{author}{\bibfnamefont{B.}~\bibnamefont{King}} \bibnamefont{and}
  \bibinfo{author}{\bibfnamefont{H.}~\bibnamefont{Ruhl}},
  \bibinfo{journal}{Phys. Rev. D} \textbf{\bibinfo{volume}{88}},
  \bibinfo{pages}{013005} (\bibinfo{year}{2013}).

\bibitem[{\citenamefont{King and Fedotov}(2018)}]{king18c}
\bibinfo{author}{\bibfnamefont{B.}~\bibnamefont{King}} \bibnamefont{and}
  \bibinfo{author}{\bibfnamefont{A.~M.} \bibnamefont{Fedotov}},
  \bibinfo{journal}{Phys. Rev.} \textbf{\bibinfo{volume}{D98}},
  \bibinfo{pages}{016005} (\bibinfo{year}{2018}), \eprint{1801.07300}.

\bibitem[{\citenamefont{Novak and Kholodov}(2012)}]{Novak:2012zz}
\bibinfo{author}{\bibfnamefont{O.~P.} \bibnamefont{Novak}} \bibnamefont{and}
  \bibinfo{author}{\bibfnamefont{R.~I.} \bibnamefont{Kholodov}},
  \bibinfo{journal}{Phys. Rev. D} \textbf{\bibinfo{volume}{86}},
  \bibinfo{pages}{105013} (\bibinfo{year}{2012}), \eprint{1210.6189}.

\bibitem[{\citenamefont{Torgrimsson}(2020)}]{Torgrimsson:2020wlz}
\bibinfo{author}{\bibfnamefont{G.}~\bibnamefont{Torgrimsson}},
  \bibinfo{journal}{Phys. Rev. D} \textbf{\bibinfo{volume}{102}},
  \bibinfo{pages}{096008} (\bibinfo{year}{2020}), \eprint{2007.08492}.

\bibitem[{\citenamefont{Titov et~al.}(2021)\citenamefont{Titov, Acosta, and
  Kampfer}}]{Titov:2021kbj}
\bibinfo{author}{\bibfnamefont{A.~I.} \bibnamefont{Titov}},
  \bibinfo{author}{\bibfnamefont{U.~H.} \bibnamefont{Acosta}},
  \bibnamefont{and} \bibinfo{author}{\bibfnamefont{B.}~\bibnamefont{Kampfer}},
  \bibinfo{journal}{Phys. Rev. A} \textbf{\bibinfo{volume}{104}},
  \bibinfo{pages}{062811} (\bibinfo{year}{2021}), \eprint{2108.13043}.

\bibitem[{\citenamefont{Kami\'nski and Krajewska}(2022)}]{Kaminski:2022uoi}
\bibinfo{author}{\bibfnamefont{J.~Z.} \bibnamefont{Kami\'nski}}
  \bibnamefont{and} \bibinfo{author}{\bibfnamefont{K.}~\bibnamefont{Krajewska}}
  (\bibinfo{year}{2022}), \eprint{2211.04716}.

\bibitem[{\citenamefont{Cole et~al.}(2018)\citenamefont{Cole, Behm, Gerstmayr,
  Blackburn, Wood, Baird, Duff, Harvey, Ilderton, Joglekar et~al.}}]{cole18}
\bibinfo{author}{\bibfnamefont{J.~M.} \bibnamefont{Cole}},
  \bibinfo{author}{\bibfnamefont{K.~T.} \bibnamefont{Behm}},
  \bibinfo{author}{\bibfnamefont{E.}~\bibnamefont{Gerstmayr}},
  \bibinfo{author}{\bibfnamefont{T.~G.} \bibnamefont{Blackburn}},
  \bibinfo{author}{\bibfnamefont{J.~C.} \bibnamefont{Wood}},
  \bibinfo{author}{\bibfnamefont{C.~D.} \bibnamefont{Baird}},
  \bibinfo{author}{\bibfnamefont{M.~J.} \bibnamefont{Duff}},
  \bibinfo{author}{\bibfnamefont{C.}~\bibnamefont{Harvey}},
  \bibinfo{author}{\bibfnamefont{A.}~\bibnamefont{Ilderton}},
  \bibinfo{author}{\bibfnamefont{A.~S.} \bibnamefont{Joglekar}},
  \bibnamefont{et~al.}, \bibinfo{journal}{Phys. Rev. X}
  \textbf{\bibinfo{volume}{8}}, \bibinfo{pages}{011020} (\bibinfo{year}{2018}).

\bibitem[{\citenamefont{Poder et~al.}(2018)\citenamefont{Poder, Tamburini,
  Sarri, Di~Piazza, Kuschel, Baird, Behm, Bohlen, Cole, Corvan
  et~al.}}]{poder18}
\bibinfo{author}{\bibfnamefont{K.}~\bibnamefont{Poder}},
  \bibinfo{author}{\bibfnamefont{M.}~\bibnamefont{Tamburini}},
  \bibinfo{author}{\bibfnamefont{G.}~\bibnamefont{Sarri}},
  \bibinfo{author}{\bibfnamefont{A.}~\bibnamefont{Di~Piazza}},
  \bibinfo{author}{\bibfnamefont{S.}~\bibnamefont{Kuschel}},
  \bibinfo{author}{\bibfnamefont{C.~D.} \bibnamefont{Baird}},
  \bibinfo{author}{\bibfnamefont{K.}~\bibnamefont{Behm}},
  \bibinfo{author}{\bibfnamefont{S.}~\bibnamefont{Bohlen}},
  \bibinfo{author}{\bibfnamefont{J.~M.} \bibnamefont{Cole}},
  \bibinfo{author}{\bibfnamefont{D.~J.} \bibnamefont{Corvan}},
  \bibnamefont{et~al.}, \bibinfo{journal}{Phys. Rev. X}
  \textbf{\bibinfo{volume}{8}}, \bibinfo{pages}{031004} (\bibinfo{year}{2018}).

\bibitem[{\citenamefont{Di~Piazza et~al.}(2020)\citenamefont{Di~Piazza,
  Wistisen, Tamburini, and Uggerh\o{}j}}]{DiPiazza:2019vwb}
\bibinfo{author}{\bibfnamefont{A.}~\bibnamefont{Di~Piazza}},
  \bibinfo{author}{\bibfnamefont{T.~N.} \bibnamefont{Wistisen}},
  \bibinfo{author}{\bibfnamefont{M.}~\bibnamefont{Tamburini}},
  \bibnamefont{and} \bibinfo{author}{\bibfnamefont{U.~I.}
  \bibnamefont{Uggerh\o{}j}}, \bibinfo{journal}{Phys. Rev. Lett.}
  \textbf{\bibinfo{volume}{124}}, \bibinfo{pages}{044801}
  (\bibinfo{year}{2020}), \eprint{1911.04749}.

\bibitem[{\citenamefont{{Salgado} et~al.}(2021)\citenamefont{{Salgado},
  {Grafenstein}, {Golub}, {D{\"o}pp}, {Eckey}, {Hollatz}, {M{\"u}ller},
  {Seidel}, {Seipt}, {Karsch} et~al.}}]{2021NJPh...23j5002S}
\bibinfo{author}{\bibfnamefont{F.~C.} \bibnamefont{{Salgado}}},
  \bibinfo{author}{\bibfnamefont{K.}~\bibnamefont{{Grafenstein}}},
  \bibinfo{author}{\bibfnamefont{A.}~\bibnamefont{{Golub}}},
  \bibinfo{author}{\bibfnamefont{A.}~\bibnamefont{{D{\"o}pp}}},
  \bibinfo{author}{\bibfnamefont{A.}~\bibnamefont{{Eckey}}},
  \bibinfo{author}{\bibfnamefont{D.}~\bibnamefont{{Hollatz}}},
  \bibinfo{author}{\bibfnamefont{C.}~\bibnamefont{{M{\"u}ller}}},
  \bibinfo{author}{\bibfnamefont{A.}~\bibnamefont{{Seidel}}},
  \bibinfo{author}{\bibfnamefont{D.}~\bibnamefont{{Seipt}}},
  \bibinfo{author}{\bibfnamefont{S.}~\bibnamefont{{Karsch}}},
  \bibnamefont{et~al.}, \bibinfo{journal}{New Journal of Physics}
  \textbf{\bibinfo{volume}{23}}, \bibinfo{eid}{105002} (\bibinfo{year}{2021}).

\bibitem[{\citenamefont{King et~al.}(2013)\citenamefont{King, Elkina, and
  Ruhl}}]{king13a}
\bibinfo{author}{\bibfnamefont{B.}~\bibnamefont{King}},
  \bibinfo{author}{\bibfnamefont{N.}~\bibnamefont{Elkina}}, \bibnamefont{and}
  \bibinfo{author}{\bibfnamefont{H.}~\bibnamefont{Ruhl}},
  \bibinfo{journal}{Phys. Rev. A} \textbf{\bibinfo{volume}{87}},
  \bibinfo{pages}{042117} (\bibinfo{year}{2013}).

\bibitem[{\citenamefont{Li et~al.}(2020)\citenamefont{Li, Shaisultanov, Chen,
  Wan, Hatsagortsyan, Keitel, and Li}}]{PhysRevLett.124.014801}
\bibinfo{author}{\bibfnamefont{Y.-F.} \bibnamefont{Li}},
  \bibinfo{author}{\bibfnamefont{R.}~\bibnamefont{Shaisultanov}},
  \bibinfo{author}{\bibfnamefont{Y.-Y.} \bibnamefont{Chen}},
  \bibinfo{author}{\bibfnamefont{F.}~\bibnamefont{Wan}},
  \bibinfo{author}{\bibfnamefont{K.~Z.} \bibnamefont{Hatsagortsyan}},
  \bibinfo{author}{\bibfnamefont{C.~H.} \bibnamefont{Keitel}},
  \bibnamefont{and} \bibinfo{author}{\bibfnamefont{J.-X.} \bibnamefont{Li}},
  \bibinfo{journal}{Phys. Rev. Lett.} \textbf{\bibinfo{volume}{124}},
  \bibinfo{pages}{014801} (\bibinfo{year}{2020}),
  \urlprefix\url{https://link.aps.org/doi/10.1103/PhysRevLett.124.014801}.

\bibitem[{\citenamefont{Wan et~al.}(2020)\citenamefont{Wan, Wang, Guo, Chen,
  Shaisultanov, Xu, Hatsagortsyan, Keitel, and Li}}]{PhysRevResearch.2.032049}
\bibinfo{author}{\bibfnamefont{F.}~\bibnamefont{Wan}},
  \bibinfo{author}{\bibfnamefont{Y.}~\bibnamefont{Wang}},
  \bibinfo{author}{\bibfnamefont{R.-T.} \bibnamefont{Guo}},
  \bibinfo{author}{\bibfnamefont{Y.-Y.} \bibnamefont{Chen}},
  \bibinfo{author}{\bibfnamefont{R.}~\bibnamefont{Shaisultanov}},
  \bibinfo{author}{\bibfnamefont{Z.-F.} \bibnamefont{Xu}},
  \bibinfo{author}{\bibfnamefont{K.~Z.} \bibnamefont{Hatsagortsyan}},
  \bibinfo{author}{\bibfnamefont{C.~H.} \bibnamefont{Keitel}},
  \bibnamefont{and} \bibinfo{author}{\bibfnamefont{J.-X.} \bibnamefont{Li}},
  \bibinfo{journal}{Phys. Rev. Research} \textbf{\bibinfo{volume}{2}},
  \bibinfo{pages}{032049} (\bibinfo{year}{2020}),
  \urlprefix\url{https://link.aps.org/doi/10.1103/PhysRevResearch.2.032049}.

\bibitem[{\citenamefont{Seipt et~al.}(2021)\citenamefont{Seipt, Ridgers, Sorbo,
  and Thomas}}]{Seipt-2021}
\bibinfo{author}{\bibfnamefont{D.}~\bibnamefont{Seipt}},
  \bibinfo{author}{\bibfnamefont{C.~P.} \bibnamefont{Ridgers}},
  \bibinfo{author}{\bibfnamefont{D.~D.} \bibnamefont{Sorbo}}, \bibnamefont{and}
  \bibinfo{author}{\bibfnamefont{A.~G.~R.} \bibnamefont{Thomas}},
  \bibinfo{journal}{New Journal of Physics} \textbf{\bibinfo{volume}{23}},
  \bibinfo{pages}{053025} (\bibinfo{year}{2021}),
  \urlprefix\url{https://doi.org/10.1088/1367-2630/abf584}.

\bibitem[{\citenamefont{Seipt et~al.}(2022)\citenamefont{Seipt, Ridgers, Sorbo,
  and Thomas}}]{Seipt-2022}
\bibinfo{author}{\bibfnamefont{D.}~\bibnamefont{Seipt}},
  \bibinfo{author}{\bibfnamefont{C.~P.} \bibnamefont{Ridgers}},
  \bibinfo{author}{\bibfnamefont{D.~D.} \bibnamefont{Sorbo}}, \bibnamefont{and}
  \bibinfo{author}{\bibfnamefont{A.~G.~R.} \bibnamefont{Thomas}},
  \bibinfo{journal}{New Journal of Physics} \textbf{\bibinfo{volume}{24}},
  \bibinfo{pages}{029501} (\bibinfo{year}{2022}),
  \urlprefix\url{https://dx.doi.org/10.1088/1367-2630/ac48e9}.

\bibitem[{\citenamefont{Ritus}(1985)}]{ritus85}
\bibinfo{author}{\bibfnamefont{V.~I.} \bibnamefont{Ritus}},
  \bibinfo{journal}{J. Russ. Laser Res.} \textbf{\bibinfo{volume}{6}},
  \bibinfo{pages}{497} (\bibinfo{year}{1985}),
  \urlprefix\url{https://doi.org/10.1007/BF01120220}.

\bibitem[{\citenamefont{Seipt and King}(2020)}]{Seipt:2020diz}
\bibinfo{author}{\bibfnamefont{D.}~\bibnamefont{Seipt}} \bibnamefont{and}
  \bibinfo{author}{\bibfnamefont{B.}~\bibnamefont{King}},
  \bibinfo{journal}{Phys. Rev. A} \textbf{\bibinfo{volume}{102}},
  \bibinfo{pages}{052805} (\bibinfo{year}{2020}), \eprint{2007.11837}.

\bibitem[{\citenamefont{Fedotov et~al.}(2022)\citenamefont{Fedotov, Ilderton,
  Karbstein, King, Seipt, Taya, and Torgrimsson}}]{Fedotov:2022ely}
\bibinfo{author}{\bibfnamefont{A.}~\bibnamefont{Fedotov}},
  \bibinfo{author}{\bibfnamefont{A.}~\bibnamefont{Ilderton}},
  \bibinfo{author}{\bibfnamefont{F.}~\bibnamefont{Karbstein}},
  \bibinfo{author}{\bibfnamefont{B.}~\bibnamefont{King}},
  \bibinfo{author}{\bibfnamefont{D.}~\bibnamefont{Seipt}},
  \bibinfo{author}{\bibfnamefont{H.}~\bibnamefont{Taya}}, \bibnamefont{and}
  \bibinfo{author}{\bibfnamefont{G.}~\bibnamefont{Torgrimsson}}
  (\bibinfo{year}{2022}), \eprint{2203.00019}.

\bibitem[{\citenamefont{Harvey et~al.}(2015)\citenamefont{Harvey, Ilderton, and
  King}}]{harvey15}
\bibinfo{author}{\bibfnamefont{C.~N.} \bibnamefont{Harvey}},
  \bibinfo{author}{\bibfnamefont{A.}~\bibnamefont{Ilderton}}, \bibnamefont{and}
  \bibinfo{author}{\bibfnamefont{B.}~\bibnamefont{King}},
  \bibinfo{journal}{Phys. Rev. A} \textbf{\bibinfo{volume}{91}},
  \bibinfo{pages}{013822} (\bibinfo{year}{2015}).

\bibitem[{\citenamefont{Di~Piazza et~al.}(2018)\citenamefont{Di~Piazza,
  Tamburini, Meuren, and Keitel}}]{DiPiazza:2017raw}
\bibinfo{author}{\bibfnamefont{A.}~\bibnamefont{Di~Piazza}},
  \bibinfo{author}{\bibfnamefont{M.}~\bibnamefont{Tamburini}},
  \bibinfo{author}{\bibfnamefont{S.}~\bibnamefont{Meuren}}, \bibnamefont{and}
  \bibinfo{author}{\bibfnamefont{C.~H.} \bibnamefont{Keitel}},
  \bibinfo{journal}{Phys. Rev. A} \textbf{\bibinfo{volume}{98}},
  \bibinfo{pages}{012134} (\bibinfo{year}{2018}), \eprint{1708.08276}.

\bibitem[{\citenamefont{Ilderton et~al.}(2019)\citenamefont{Ilderton, King, and
  Seipt}}]{Ilderton:2018nws}
\bibinfo{author}{\bibfnamefont{A.}~\bibnamefont{Ilderton}},
  \bibinfo{author}{\bibfnamefont{B.}~\bibnamefont{King}}, \bibnamefont{and}
  \bibinfo{author}{\bibfnamefont{D.}~\bibnamefont{Seipt}},
  \bibinfo{journal}{Phys. Rev. A} \textbf{\bibinfo{volume}{99}},
  \bibinfo{pages}{042121} (\bibinfo{year}{2019}), \eprint{1808.10339}.

\bibitem[{\citenamefont{Di~Piazza et~al.}(2019)\citenamefont{Di~Piazza,
  Tamburini, Meuren, and Keitel}}]{DiPiazza:2018bfu}
\bibinfo{author}{\bibfnamefont{A.}~\bibnamefont{Di~Piazza}},
  \bibinfo{author}{\bibfnamefont{M.}~\bibnamefont{Tamburini}},
  \bibinfo{author}{\bibfnamefont{S.}~\bibnamefont{Meuren}}, \bibnamefont{and}
  \bibinfo{author}{\bibfnamefont{C.~H.} \bibnamefont{Keitel}},
  \bibinfo{journal}{Phys. Rev. A} \textbf{\bibinfo{volume}{99}},
  \bibinfo{pages}{022125} (\bibinfo{year}{2019}), \eprint{1811.05834}.

\bibitem[{\citenamefont{King}(2020)}]{King:2019igt}
\bibinfo{author}{\bibfnamefont{B.}~\bibnamefont{King}}, \bibinfo{journal}{Phys.
  Rev. A} \textbf{\bibinfo{volume}{101}}, \bibinfo{pages}{042508}
  (\bibinfo{year}{2020}), \eprint{1908.06985}.

\bibitem[{\citenamefont{Heinzl et~al.}(2020)\citenamefont{Heinzl, King, and
  MacLeod}}]{Heinzl:2020ynb}
\bibinfo{author}{\bibfnamefont{T.}~\bibnamefont{Heinzl}},
  \bibinfo{author}{\bibfnamefont{B.}~\bibnamefont{King}}, \bibnamefont{and}
  \bibinfo{author}{\bibfnamefont{A.~J.} \bibnamefont{MacLeod}},
  \bibinfo{journal}{Phys. Rev. A} \textbf{\bibinfo{volume}{102}},
  \bibinfo{pages}{063110} (\bibinfo{year}{2020}),
  \urlprefix\url{https://link.aps.org/doi/10.1103/PhysRevA.102.063110}.

\bibitem[{\citenamefont{Blackburn and King}(2022)}]{Blackburn:2021cuq}
\bibinfo{author}{\bibfnamefont{T.~G.} \bibnamefont{Blackburn}}
  \bibnamefont{and} \bibinfo{author}{\bibfnamefont{B.}~\bibnamefont{King}},
  \bibinfo{journal}{Eur. Phys. J. C} \textbf{\bibinfo{volume}{82}},
  \bibinfo{pages}{44} (\bibinfo{year}{2022}), \eprint{2108.10883}.

\bibitem[{\citenamefont{Blackburn et~al.}(2021)\citenamefont{Blackburn,
  MacLeod, and King}}]{Blackburn:2021rqm}
\bibinfo{author}{\bibfnamefont{T.~G.} \bibnamefont{Blackburn}},
  \bibinfo{author}{\bibfnamefont{A.~J.} \bibnamefont{MacLeod}},
  \bibnamefont{and} \bibinfo{author}{\bibfnamefont{B.}~\bibnamefont{King}},
  \bibinfo{journal}{New J. Phys.} \textbf{\bibinfo{volume}{23}},
  \bibinfo{pages}{085008} (\bibinfo{year}{2021}), \eprint{2103.06673}.

\bibitem[{\citenamefont{Turner et~al.}(2022)\citenamefont{Turner, Bulanov,
  Benedetti, Gonsalves, Leemans, Nakamura, van Tilborg, Schroeder, Geddes, and
  Esarey}}]{Turner:2022hch}
\bibinfo{author}{\bibfnamefont{M.}~\bibnamefont{Turner}},
  \bibinfo{author}{\bibfnamefont{S.~S.} \bibnamefont{Bulanov}},
  \bibinfo{author}{\bibfnamefont{C.}~\bibnamefont{Benedetti}},
  \bibinfo{author}{\bibfnamefont{A.~J.} \bibnamefont{Gonsalves}},
  \bibinfo{author}{\bibfnamefont{W.~P.} \bibnamefont{Leemans}},
  \bibinfo{author}{\bibfnamefont{K.}~\bibnamefont{Nakamura}},
  \bibinfo{author}{\bibfnamefont{J.}~\bibnamefont{van Tilborg}},
  \bibinfo{author}{\bibfnamefont{C.~B.} \bibnamefont{Schroeder}},
  \bibinfo{author}{\bibfnamefont{C.~G.~R.} \bibnamefont{Geddes}},
  \bibnamefont{and} \bibinfo{author}{\bibfnamefont{E.}~\bibnamefont{Esarey}}
  (\bibinfo{year}{2022}), \eprint{2210.09214}.

\bibitem[{\citenamefont{Yokoya and Chen}(1992)}]{Yokoya:1991qz}
\bibinfo{author}{\bibfnamefont{K.}~\bibnamefont{Yokoya}} \bibnamefont{and}
  \bibinfo{author}{\bibfnamefont{P.}~\bibnamefont{Chen}},
  \bibinfo{journal}{Lect. Notes Phys.} \textbf{\bibinfo{volume}{400}},
  \bibinfo{pages}{415} (\bibinfo{year}{1992}).

\bibitem[{\citenamefont{Hartin}(2018)}]{Hartin:2018egj}
\bibinfo{author}{\bibfnamefont{A.}~\bibnamefont{Hartin}},
  \bibinfo{journal}{Int. J. Mod. Phys. A} \textbf{\bibinfo{volume}{33}},
  \bibinfo{pages}{1830011} (\bibinfo{year}{2018}), \eprint{1804.02934}.

\bibitem[{\citenamefont{Ivanov et~al.}(2004)\citenamefont{Ivanov, Kotkin, and
  Serbo}}]{Ivanov:2004fi}
\bibinfo{author}{\bibfnamefont{D.~Y.} \bibnamefont{Ivanov}},
  \bibinfo{author}{\bibfnamefont{G.~L.} \bibnamefont{Kotkin}},
  \bibnamefont{and} \bibinfo{author}{\bibfnamefont{V.~G.} \bibnamefont{Serbo}},
  \bibinfo{journal}{Eur. Phys. J. C} \textbf{\bibinfo{volume}{36}},
  \bibinfo{pages}{127} (\bibinfo{year}{2004}), \eprint{hep-ph/0402139}.

\bibitem[{\citenamefont{King and Tang}(2020)}]{TangPRA022809}
\bibinfo{author}{\bibfnamefont{B.}~\bibnamefont{King}} \bibnamefont{and}
  \bibinfo{author}{\bibfnamefont{S.}~\bibnamefont{Tang}},
  \bibinfo{journal}{Phys. Rev. A} \textbf{\bibinfo{volume}{102}},
  \bibinfo{pages}{022809} (\bibinfo{year}{2020}),
  \urlprefix\url{https://link.aps.org/doi/10.1103/PhysRevA.102.022809}.

\bibitem[{\citenamefont{Tang et~al.}(2020)\citenamefont{Tang, King, and
  Hu}}]{TANG2020135701}
\bibinfo{author}{\bibfnamefont{S.}~\bibnamefont{Tang}},
  \bibinfo{author}{\bibfnamefont{B.}~\bibnamefont{King}}, \bibnamefont{and}
  \bibinfo{author}{\bibfnamefont{H.}~\bibnamefont{Hu}},
  \bibinfo{journal}{Physics Letters B} \textbf{\bibinfo{volume}{809}},
  \bibinfo{pages}{135701} (\bibinfo{year}{2020}), ISSN
  \bibinfo{issn}{0370-2693},
  \urlprefix\url{https://www.sciencedirect.com/science/article/pii/S0370269320305049}.

\bibitem[{\citenamefont{Tang}(2022)}]{Tang:2022a}
\bibinfo{author}{\bibfnamefont{S.}~\bibnamefont{Tang}}, \bibinfo{journal}{Phys.
  Rev. D} \textbf{\bibinfo{volume}{105}}, \bibinfo{pages}{056018}
  (\bibinfo{year}{2022}),
  \urlprefix\url{https://link.aps.org/doi/10.1103/PhysRevD.105.056018}.

\bibitem[{\citenamefont{Gao and Tang}(2022)}]{TangPRD056003}
\bibinfo{author}{\bibfnamefont{Y.}~\bibnamefont{Gao}} \bibnamefont{and}
  \bibinfo{author}{\bibfnamefont{S.}~\bibnamefont{Tang}},
  \bibinfo{journal}{Phys. Rev. D} \textbf{\bibinfo{volume}{106}},
  \bibinfo{pages}{056003} (\bibinfo{year}{2022}),
  \urlprefix\url{https://link.aps.org/doi/10.1103/PhysRevD.106.056003}.

\bibitem[{\citenamefont{King}(2021)}]{King:2020hsk}
\bibinfo{author}{\bibfnamefont{B.}~\bibnamefont{King}}, \bibinfo{journal}{Phys.
  Rev. D} \textbf{\bibinfo{volume}{103}}, \bibinfo{pages}{036018}
  (\bibinfo{year}{2021}), \eprint{2012.05920}.

\bibitem[{\citenamefont{Tang and King}(2021)}]{Tang:2021qht}
\bibinfo{author}{\bibfnamefont{S.}~\bibnamefont{Tang}} \bibnamefont{and}
  \bibinfo{author}{\bibfnamefont{B.}~\bibnamefont{King}},
  \bibinfo{journal}{Phys. Rev. D} \textbf{\bibinfo{volume}{104}},
  \bibinfo{pages}{096019} (\bibinfo{year}{2021}), \eprint{2109.00555}.

\bibitem[{\citenamefont{Di~Piazza et~al.}(2012)\citenamefont{Di~Piazza,
  M\"uller, Hatsagortsyan, and Keitel}}]{dipiazza12}
\bibinfo{author}{\bibfnamefont{A.}~\bibnamefont{Di~Piazza}},
  \bibinfo{author}{\bibfnamefont{C.}~\bibnamefont{M\"uller}},
  \bibinfo{author}{\bibfnamefont{K.~Z.} \bibnamefont{Hatsagortsyan}},
  \bibnamefont{and} \bibinfo{author}{\bibfnamefont{C.~H.}
  \bibnamefont{Keitel}}, \bibinfo{journal}{Rev. Mod. Phys.}
  \textbf{\bibinfo{volume}{84}}, \bibinfo{pages}{1177} (\bibinfo{year}{2012}),
  \urlprefix\url{https://link.aps.org/doi/10.1103/RevModPhys.84.1177}.

\bibitem[{\citenamefont{Gonoskov et~al.}(2022)\citenamefont{Gonoskov,
  Blackburn, Marklund, and Bulanov}}]{RMP2022_045001}
\bibinfo{author}{\bibfnamefont{A.}~\bibnamefont{Gonoskov}},
  \bibinfo{author}{\bibfnamefont{T.~G.} \bibnamefont{Blackburn}},
  \bibinfo{author}{\bibfnamefont{M.}~\bibnamefont{Marklund}}, \bibnamefont{and}
  \bibinfo{author}{\bibfnamefont{S.~S.} \bibnamefont{Bulanov}},
  \bibinfo{journal}{Rev. Mod. Phys.} \textbf{\bibinfo{volume}{94}},
  \bibinfo{pages}{045001} (\bibinfo{year}{2022}),
  \urlprefix\url{https://link.aps.org/doi/10.1103/RevModPhys.94.045001}.

\bibitem[{\citenamefont{Bell and Kirk}(2008)}]{PRL2008200403}
\bibinfo{author}{\bibfnamefont{A.~R.} \bibnamefont{Bell}} \bibnamefont{and}
  \bibinfo{author}{\bibfnamefont{J.~G.} \bibnamefont{Kirk}},
  \bibinfo{journal}{Phys. Rev. Lett.} \textbf{\bibinfo{volume}{101}},
  \bibinfo{pages}{200403} (\bibinfo{year}{2008}),
  \urlprefix\url{https://link.aps.org/doi/10.1103/PhysRevLett.101.200403}.

\bibitem[{\citenamefont{Kirk et~al.}(2009)\citenamefont{Kirk, Bell, and
  Arka}}]{Kirk_2009}
\bibinfo{author}{\bibfnamefont{J.~G.} \bibnamefont{Kirk}},
  \bibinfo{author}{\bibfnamefont{A.~R.} \bibnamefont{Bell}}, \bibnamefont{and}
  \bibinfo{author}{\bibfnamefont{I.}~\bibnamefont{Arka}},
  \bibinfo{journal}{Plasma Physics and Controlled Fusion}
  \textbf{\bibinfo{volume}{51}}, \bibinfo{pages}{085008}
  (\bibinfo{year}{2009}),
  \urlprefix\url{https://dx.doi.org/10.1088/0741-3335/51/8/085008}.

\bibitem[{\citenamefont{Nerush et~al.}(2011)}]{nerush11}
\bibinfo{author}{\bibfnamefont{E.~N.} \bibnamefont{Nerush}}
  \bibnamefont{et~al.}, \bibinfo{journal}{Phys. Rev. Lett.}
  \textbf{\bibinfo{volume}{106}}, \bibinfo{pages}{035001}
  (\bibinfo{year}{2011}).

\bibitem[{\citenamefont{Elkina et~al.}(2011)}]{elkina11}
\bibinfo{author}{\bibfnamefont{N.~V.} \bibnamefont{Elkina}}
  \bibnamefont{et~al.}, \bibinfo{journal}{Phys. Rev. ST Accel. Beams}
  \textbf{\bibinfo{volume}{14}}, \bibinfo{pages}{054401}
  (\bibinfo{year}{2011}).

\bibitem[{\citenamefont{Tang et~al.}(2014)\citenamefont{Tang, Bake, Wang, and
  Xie}}]{TangPRA2014}
\bibinfo{author}{\bibfnamefont{S.}~\bibnamefont{Tang}},
  \bibinfo{author}{\bibfnamefont{M.~A.} \bibnamefont{Bake}},
  \bibinfo{author}{\bibfnamefont{H.-Y.} \bibnamefont{Wang}}, \bibnamefont{and}
  \bibinfo{author}{\bibfnamefont{B.-S.} \bibnamefont{Xie}},
  \bibinfo{journal}{Phys. Rev. A} \textbf{\bibinfo{volume}{89}},
  \bibinfo{pages}{022105} (\bibinfo{year}{2014}),
  \urlprefix\url{https://link.aps.org/doi/10.1103/PhysRevA.89.022105}.

\bibitem[{\citenamefont{Gonoskov et~al.}(2017)\citenamefont{Gonoskov, Bashinov,
  Bastrakov, Efimenko, Ilderton, Kim, Marklund, Meyerov, Muraviev, and
  Sergeev}}]{gonoskov17}
\bibinfo{author}{\bibfnamefont{A.}~\bibnamefont{Gonoskov}},
  \bibinfo{author}{\bibfnamefont{A.}~\bibnamefont{Bashinov}},
  \bibinfo{author}{\bibfnamefont{S.}~\bibnamefont{Bastrakov}},
  \bibinfo{author}{\bibfnamefont{E.}~\bibnamefont{Efimenko}},
  \bibinfo{author}{\bibfnamefont{A.}~\bibnamefont{Ilderton}},
  \bibinfo{author}{\bibfnamefont{A.}~\bibnamefont{Kim}},
  \bibinfo{author}{\bibfnamefont{M.}~\bibnamefont{Marklund}},
  \bibinfo{author}{\bibfnamefont{I.}~\bibnamefont{Meyerov}},
  \bibinfo{author}{\bibfnamefont{A.}~\bibnamefont{Muraviev}}, \bibnamefont{and}
  \bibinfo{author}{\bibfnamefont{A.}~\bibnamefont{Sergeev}},
  \bibinfo{journal}{Phys. Rev. X} \textbf{\bibinfo{volume}{7}},
  \bibinfo{pages}{041003} (\bibinfo{year}{2017}),
  \urlprefix\url{https://link.aps.org/doi/10.1103/PhysRevX.7.041003}.

\bibitem[{\citenamefont{Samsonov et~al.}(2019)\citenamefont{Samsonov, Nerush,
  and Kostyukov}}]{Samsonov:2018nff}
\bibinfo{author}{\bibfnamefont{A.~S.} \bibnamefont{Samsonov}},
  \bibinfo{author}{\bibfnamefont{E.~N.} \bibnamefont{Nerush}},
  \bibnamefont{and} \bibinfo{author}{\bibfnamefont{I.~Y.}
  \bibnamefont{Kostyukov}}, \bibinfo{journal}{Sci. Rep.}
  \textbf{\bibinfo{volume}{9}}, \bibinfo{pages}{11133} (\bibinfo{year}{2019}),
  \eprint{1809.06115}.

\bibitem[{\citenamefont{Sampath and Tamburini}(2018)}]{Khudik:2018hkr}
\bibinfo{author}{\bibfnamefont{A.}~\bibnamefont{Sampath}} \bibnamefont{and}
  \bibinfo{author}{\bibfnamefont{M.}~\bibnamefont{Tamburini}},
  \bibinfo{journal}{Phys. Plasmas} \textbf{\bibinfo{volume}{25}},
  \bibinfo{pages}{083104} (\bibinfo{year}{2018}), \eprint{1807.09747}.

\bibitem[{\citenamefont{Torgrimsson}(2021)}]{Torgrimsson:2020gws}
\bibinfo{author}{\bibfnamefont{G.}~\bibnamefont{Torgrimsson}},
  \bibinfo{journal}{New J. Phys.} \textbf{\bibinfo{volume}{23}},
  \bibinfo{pages}{065001} (\bibinfo{year}{2021}), \eprint{2012.12701}.

\bibitem[{\citenamefont{Harvey et~al.}(2009)\citenamefont{Harvey, Heinzl, and
  Ilderton}}]{Harvey09}
\bibinfo{author}{\bibfnamefont{C.}~\bibnamefont{Harvey}},
  \bibinfo{author}{\bibfnamefont{T.}~\bibnamefont{Heinzl}}, \bibnamefont{and}
  \bibinfo{author}{\bibfnamefont{A.}~\bibnamefont{Ilderton}},
  \bibinfo{journal}{Phys. Rev. A} \textbf{\bibinfo{volume}{79}},
  \bibinfo{pages}{063407} (\bibinfo{year}{2009}).

\bibitem[{\citenamefont{Toll}(1952)}]{Toll:1952rq}
\bibinfo{author}{\bibfnamefont{J.~S.} \bibnamefont{Toll}}, Ph.D. thesis,
  \bibinfo{school}{Princeton U.} (\bibinfo{year}{1952}).

\bibitem[{\citenamefont{Korsch et~al.}(2006)\citenamefont{Korsch, Klumpp, and
  Witthaut}}]{Korsch_2006}
\bibinfo{author}{\bibfnamefont{H.~J.} \bibnamefont{Korsch}},
  \bibinfo{author}{\bibfnamefont{A.}~\bibnamefont{Klumpp}}, \bibnamefont{and}
  \bibinfo{author}{\bibfnamefont{D.}~\bibnamefont{Witthaut}},
  \bibinfo{journal}{Journal of Physics A: Mathematical and General}
  \textbf{\bibinfo{volume}{39}}, \bibinfo{pages}{14947} (\bibinfo{year}{2006}),
  \urlprefix\url{https://dx.doi.org/10.1088/0305-4470/39/48/008}.

\bibitem[{\citenamefont{L\"otstedt and Jentschura}(2009)}]{PhysRevE.79.026707}
\bibinfo{author}{\bibfnamefont{E.}~\bibnamefont{L\"otstedt}} \bibnamefont{and}
  \bibinfo{author}{\bibfnamefont{U.~D.} \bibnamefont{Jentschura}},
  \bibinfo{journal}{Phys. Rev. E} \textbf{\bibinfo{volume}{79}},
  \bibinfo{pages}{026707} (\bibinfo{year}{2009}),
  \urlprefix\url{https://link.aps.org/doi/10.1103/PhysRevE.79.026707}.

\bibitem[{\citenamefont{Titov et~al.}(2012)\citenamefont{Titov, Takabe,
  K\"ampfer, and Hosaka}}]{PhysRevLett.108.240406}
\bibinfo{author}{\bibfnamefont{A.~I.} \bibnamefont{Titov}},
  \bibinfo{author}{\bibfnamefont{H.}~\bibnamefont{Takabe}},
  \bibinfo{author}{\bibfnamefont{B.}~\bibnamefont{K\"ampfer}},
  \bibnamefont{and} \bibinfo{author}{\bibfnamefont{A.}~\bibnamefont{Hosaka}},
  \bibinfo{journal}{Phys. Rev. Lett.} \textbf{\bibinfo{volume}{108}},
  \bibinfo{pages}{240406} (\bibinfo{year}{2012}),
  \urlprefix\url{https://link.aps.org/doi/10.1103/PhysRevLett.108.240406}.

\bibitem[{\citenamefont{Nousch et~al.}(2012)\citenamefont{Nousch, Seipt,
  K\"ampfer, and Titov}}]{seipt12b}
\bibinfo{author}{\bibfnamefont{T.}~\bibnamefont{Nousch}},
  \bibinfo{author}{\bibfnamefont{D.}~\bibnamefont{Seipt}},
  \bibinfo{author}{\bibfnamefont{B.}~\bibnamefont{K\"ampfer}},
  \bibnamefont{and} \bibinfo{author}{\bibfnamefont{A.}~\bibnamefont{Titov}},
  \bibinfo{journal}{Physics Letters B} \textbf{\bibinfo{volume}{715}},
  \bibinfo{pages}{246} (\bibinfo{year}{2012}), ISSN \bibinfo{issn}{0370-2693},
  \urlprefix\url{https://www.sciencedirect.com/science/article/pii/S0370269312007800}.

\end{thebibliography}

\end{document}